\documentstyle[12pt]{article}

\advance\voffset by -1.3cm
\advance\hoffset by -1.5cm
\def\be{\begin{equation}}
\def\ee{\end{equation}}
\def\Cop{\bbbc}
\def\Zop{\bbbz}
\def\Rop{\bbbr}
\def\Nop{\bbbn}
\def\bbbz {{\sf Z\!\!Z}}
\def\bbbr {{\rm I\!R}}
\def\bbbn {{\rm I\!N}}
\def\bbbc{{\mathchoice {\setbox0=\hbox{$\displaystyle\rm C$}\hbox{\hbox
to0pt{\kern0.4\wd0\vrule height0.9\ht0\hss}\box0}}
{\setbox0=\hbox{$\textstyle\rm C$}\hbox{\hbox
to0pt{\kern0.4\wd0\vrule height0.9\ht0\hss}\box0}}
{\setbox0=\hbox{$\scriptstyle\rm C$}\hbox{\hbox
to0pt{\kern0.4\wd0\vrule height0.9\ht0\hss}\box0}}
{\setbox0=\hbox{$\scriptscriptstyle\rm C$}\hbox{\hbox
to0pt{\kern0.4\wd0\vrule height0.9\ht0\hss}\box0}}}}

\def\pmb#1{\setbox0=\hbox{#1}%
 \kern-.025em\copy0\kern-\wd0
 \kern.05em\copy0\kern-\wd0
 \kern-.025em\raise.0433em\box0 }
\def\sq{\hbox{\rlap{$\sqcap$}$\sqcup$}}
\def\qed{\ifmmode\sq\else{\unskip\nobreak\hfil
\penalty50\hskip1em\null\nobreak\hfil\sq
\parfillskip=0pt\finalhyphendemerits=0\endgraf}\fi}
\def\noi {\noindent}

\def\A{{\cal A}}
\def\C{{\cal C}}
\def\E{{\cal E}}
\def\P{{\cal P}}
\def\U{{\cal U}}
\def\a{{\bf a}}
\def\bOmega{{\bf \Omega}}
\def\bQ{{\bf Q}}

\def\bB{{\bf B}}
\def\bj{{\bf j}}
\def\H{{\cal H}}
\def\R{{\cal R}}
\def\M{{\cal M}}

\def\B{{\cal B}}
\def\U{{\cal U}}

\def\G{{\cal G}}

\def\F{{\cal F}}

\def\tr{{\rm tr}}
\def\o{{\otimes}}
\def\vare{{\varepsilon}}
\def\bvare{{\pmb{$\vare$}}}
\def\llangle{{\langle\langle}}
\def\rrangle{{\rangle\rangle}}
\def\bbbone {{\mathchoice {{\rm 1\mskip-4mu l}} {{\rm 1\mskip-4mu l}}
{{\rm 1\mskip-4.5mu l}} {{\rm 1\mskip-5mu l}}}}

\def\half {\frac{1}{2}}

\newcommand{\ket}[1]{|#1\rangle}
\newcommand{\bra}[1]{\langle#1|}

\begin{document}

\thispagestyle{empty}
\def\thefootnote{\fnsymbol{footnote}}
\begin{flushright}
  HUTP-97/A018 \\
  MIT-CTP-2631 \\
  hep-th/9705038
\end{flushright} \vskip 1.0cm

\begin{center}\LARGE
{\bf Tensor Constructions of Open  String Theories I: Foundations}
\end{center} \vskip 1.0cm
\begin{center}\large
       Matthias R.\ Gaberdiel%
       \footnote{E-mail  address: {\tt gaberd@string.harvard.edu}}
       and 
       Barton Zwiebach%
       \footnote{E-mail  address: {\tt zwiebach@irene.mit.edu}}
       \end{center}
\vskip0.5cm

\begin{center}
Lyman Laboratory of Physics\\
Harvard University\\
Cambridge, MA 02138
\end{center}
\vskip 1em
\begin{center}
May 1997
\end{center}
\vskip 1cm
\begin{abstract}
The possible tensor constructions of open string theories are analyzed
from first principles. To this end the algebraic framework of open
string field theory is clarified, including the role of the
homotopy associative $A_\infty$ algebra, the odd symplectic
structure, cyclicity, star conjugation, and twist. It is also shown
that two string theories are off-shell equivalent if the corresponding
homotopy associative algebras are homotopy equivalent in a strict
sense. 

It is demonstrated that a homotopy associative star algebra with a
compatible even bilinear form can be attached to an open string theory.
If this algebra does not have a spacetime interpretation,
positivity and the existence of a conserved ghost number require  
that its cohomology  is at degree zero, and that it  
has the structure of a direct sum of full matrix
algebras. The resulting string theory is shown to be  
physically equivalent to a string theory with a familiar open string
gauge group.
\end{abstract}

\vfill

\setcounter{footnote}{0}
\def\thefootnote{\arabic{footnote}}
\newpage

\setcounter{page}{1}

\tableofcontents

\renewcommand{\theequation}{\thesection.\arabic{equation}}

\section{Introduction}
\setcounter{equation}{0}

The construction of internal symmetries of open string theories was
initiated by Chan and Paton \cite{chanpaton}, who observed that it is
consistent to associate an internal matrix to each open string. This
construction leads to open string theories whose gauge group is
$U(n)$. Some time later it was realized by Schwarz \cite{schwarz} that
additional constructions are possible if the orientation reversal
symmetry of the underlying open string theory is used; the resulting
string theory can then have the gauge group $SO(n)$ or
$USp(2n)$. Shortly afterwards it was shown by Marcus and Sagnotti
\cite{MarSag} that these constructions are in essence the only
possible ones. Their argument relies on the fact that the
factorization property of the string amplitudes implies certain
identities for the various products of 
internal matrices. It is then argued that this requires the space of
internal matrices to be the real form of a semisimple complex algebra
with unit. Because of Wedderburn's structure theorem, the general
structure of these semisimple algebras is known, and the result
follows.

While this was a very effective method to discuss the possible
tensor constructions
at the time, later developments have shown that
more sophisticated constructions are possible (see {\it e.g.} the
constructions of \cite{BSag, BG, GimPol}). Moreover, with string field
theory now being better understood, it is also feasible to give a more
conceptual analysis of the possible 
tensor constructions.
Such an analysis can be considerably more general and systematic.
\medskip

We say that a  string theory theory is consistent if it is defined by
an Grassmann even action $S$, and a
Batalin-Vilkovisky (BV)  antibracket $\{ \cdot , \cdot \}$ satisfying
the conditions
\begin{list}{(\roman{enumi})}{\usecounter{enumi}}
\item $S$  satisfies the classical\footnote{In this paper we shall
only analyze tree-level constraints.} master
equation  $\{S,S\}=0$. 
\item  $S$ is real.
\item The kinetic term of $S$ defines a positive definite form
on physical states.   
\end{list}

The first condition, as is well-known 
\cite{batalinvilkovisky,henneauxteitelboim},
implies gauge invariance of the action,  and independence of the
S-matrix from gauge fixing conditions. 
The reality of the action guarantees that the Hamiltonian is
hermitian, and condition (iii) means that the physical states have
positive norms. This last condition is somewhat delicate, as the
positive definite form only arises after gauge fixing; this will be
explained in some detail in section~4.1.  The factorization
property of the amplitudes need not be imposed as a separate 
condition, as it is guaranteed by the fact that we are dealing
with a string field theory.

It was realized by Witten \cite{Witten} that the algebraic structure
of an open string field theory can be taken to be that
of a differential graded associative algebra endowed with a
trace. Such a structure allows the construction of an action that
satisfies the classical BV master equation, and
thus admits consistent quantization (see \cite{Thorn}). 
It became apparent later 
that strict associativity is not
essential to the construction. Indeed, the covariantized light-cone open
string theory of Ref.\cite{hikko}, which is consistent at tree level,
has a quartic interaction as a consequence of the lack of associativity 
of the light-cone style open string product.  
After closed string field theory was formulated and the role of
homotopy Lie algebras ($L_\infty$ algebras) was recognized \cite{Z1},
it became clear  that the  factorizable  theory of open and closed
strings \cite{Z2}, requires the  framework of
a differential graded algebra that {\it is not strictly
associative}, but rather homotopy associative.\footnote{This is clear
from the fact that the classical open string sector has interactions
to all orders.} Such 
algebras are usually referred to as $A_\infty$ algebras, and were
originally introduced by Stasheff \cite{Stash}. 
We shall therefore discuss the axioms of open string field theory in
the language of $A_{\infty}$ algebras; this contains trivially the
associative case as well. We shall explain that 
the constraints (i--iii) mentioned above can be satisfied provided the
theory has the following {\it basic algebraic structure}:
\begin{list}{(\roman{enumi}')}{\usecounter{enumi}}
\item An $A_\infty$ algebra $\A$ endowed with an 
odd, invertible symplectic form. 
This bilinear form must satisfy certain cyclicity
conditions.\footnote{This is similar but not
quite as strong as demanding the existence of a trace in the
algebra.} 
\item The algebra $\A$ is equipped with a compatible star involution.
\item There exists a positive definite sesquilinear form on the
physical states which is associated to the symplectic form
and the star involution.
\end{list}
The $A_\infty$ algebra will be assumed to be $\Zop$ graded. This
degree is associated to the ghost number, and it guarantees that a
conserved target space ghost number exists. 

We shall assume that the original open string theory has this 
basic algebraic structure, and we shall analyze under what conditions
the same will be true for a suitably defined tensor theory.

In a first step we will show that no generality can be gained by
assuming that the set $\a$ which is to be attached to the open string
theory has less structure than that of a complex vector space.
Indeed, as the underlying open string theory
forms already a complex vector space $\A$, we can replace $\a$
by a suitable vector space without changing the resulting
construction.\footnote{For example, if this set is a group
$G$, we can replace it by (the vector space of) its group algebra
$\Cop G$.} We are thus led to consider the (vector space) tensor
product $\A \o \a$. In order for it to carry the structure of an
$A_\infty$ algebra, we explain that it is sufficient to assume that the
vector space $\a$ also carries the structure of a ($\Zop$ graded)
$A_\infty$ algebra; the construction of the $A_\infty$ structure on
the tensor product is nontrivial and seems not to have been considered
before in the mathematical literature. Here we only build explicitly 
the differential, the product, and the first homotopy of the tensor
product algebra. When one of the algebras is associative, we give
explicit expressions for all the homotopies.

As part of the basic structure we also need an invertible, odd,
bilinear form on $\A \o \a$. Given that $\A$ has an invertible, odd
bilinear form, we need an invertible {\it even} bilinear form on $\a$.
The fact that the form is even cannot be changed by redefining the
degree of the elements of the algebra, and it therefore follows that
the sector $\a$ must be structurally different from the sector
corresponding to the underlying string theory.\footnote{The process of
gauge fixing, however, seems to induce an even form (from the odd
form), and this might give rise to interesting possibilities as will
be alluded to in the conclusion.}

The bilinear form on the tensor theory has to satisfy suitable
cyclicity properties, and this is guaranteed provided that the bilinear
form on $\a$ is cyclic. Furthermore, if $\a$ has a star-structure, we
can define a star-structure on the tensor algebra. This leaves us with
condition (iii') which turns out to be very restrictive. Since the
differential on $\A\o\a$ is the sum $(Q\o \bbbone + \bbbone \o q)$ of
the two differentials, the physical states of the tensor theory are
states of the form 
$H^{rel}(\A)\o H(\a)$, where $H^{rel}(\A)$ denotes the physical
states of the original theory (the relative cohomology classes of the
differential $Q$ of $\A$) and $H(\a)$ are the cohomology classes of
$q$ on $\a$. 
This is appropriate for the case when the degrees of freedom
described by $\a$ do not have any spacetime interpretation.
We show that positivity requires that $H(\a)$ is
concentrated at a single degree, a degree that can be taken to be
zero and corresponds to even elements of $\a$.  
In addition, the
bilinear form on $\a$ must induce a positive definite sesquilinear
form on $H(\a)$, {\it i.e.}  $\langle h^* , h \rangle > 0$ for all
nonvanishing $h\in H(\a)$.

It then follows that the three consistency conditions of the tensor
theory are satisfied provided that
\begin{list}{(\roman{enumi}')}{\usecounter{enumi}}
\item $\a$ is an $A_\infty$ algebra with an invertible even bilinear
form with  cyclicity properties.  
\item  $\a$ is equipped with a compatible star involution.
\item $H(\a)$ must be concentrated at zero degree,
and the bilinear form together with the star involution must induce a
positive definite sesquilinear form on $H(\a)$.
\end{list}

The rest of this paper is involved with analyzing
the above conditions. A fundamental result follows with
little work: {\it $H(\a)$ defines a semisimple associative algebra
over $\Cop$}. Indeed, the possibly non-associative product on $\a$
always induces an associative product on $H(\a)$. Moreover
semisimplicity follows from the existence of a positive definite form
on $H(\a)$.  Any semisimple algebra can be written as a direct sum of
a nilpotent ideal \footnote{This ideal is generated by elements whose
multiplication with any element in the algebra is trivial.  In the
tensor theory these elements correspond to additional non-interacting
copies of the free open string.} 
and a semisimple algebra with a unit. By Wedderburn's theorem, a
semisimple algebra with a unit is a direct sum of full matrix
algebras, and we therefore find that the cohomology is of the form 
\be
\label{mainreqx}
H(\a) = I \bigoplus_{i} \M_{n_i}(\Cop) \, , \quad \hbox{deg} (h) =0,
\,\, \hbox{for}\, h\in H(\a)\,, 
\ee 
where $I$ denotes the nilpotent ideal, and $\M_{n}(\Cop)$ denotes the
full matrix algebra of $n\times n$ matrices with complex entries.
\medskip

It remains to analyze which of the resulting tensor theories are
equivalent. In fact, two string field theories are strictly equivalent
({\it i.e.} equivalent off-shell) if their actions differ by a
canonical transformation of the BV antibracket. On the algebraic
level, this translates into the
property that the two homotopy associative algebras are
homotopy equivalent, where the homotopy equivalence induces an
isomorphism of vector spaces and satisfies a certain cyclicity
condition. We show further that two tensor theories are strictly
equivalent if the homotopy algebras that are attached are
homotopy equivalent, where again the homotopy equivalence induces an
isomorphism of vector spaces and satisfies a certain cyclicity
condition. This defines an equivalence 
relation on the space of all $A_\infty$ algebras which is somewhat
stronger than the usual homotopy equivalence. Indeed, it is known
that every homotopy associative algebra is homotopy equivalent to an
associative algebra \cite{Stasheff}, but
this is not true if we require the underlying vector spaces to be
isomorphic as there exist non-trivial deformations of associative
algebras to homotopy associative algebras \cite{PSch, markl}. On the
other hand, it is not clear whether there are deformations that
respect the conditions (i'-iii').

There exists, however, a weaker notion of equivalence which is 
relevant in physics, and which we shall therefore call 
{\it physical equivalence}. This notion of equivalence means that
there exists an identification of the physical degrees of freedom, and
that the tree-level S-matrices (of the physical states) agree.
We shall be able to show that the tensor theory corresponding to 
$\A\o\a$ is physically equivalent to the theory where we replace 
the $A_\infty$ algebra $\a$ by 
its cohomology $H(\a)$. This surprisingly strong result follows
essentially from the fact that $H(\a)$ is at degree zero, and that
there exists a conserved degree corresponding to the ghost number.  
As $H(\a)$ has the structure (\ref{mainreqx}), this   
implies that all the tensor theories are physically equivalent to
the familiar Chan Paton constructions, whose gauge groups are direct
products of $U(n)$ groups.  
\medskip

More general constructions are possible if the open string theory has
extra structure beyond the basic structure discussed before. For
example, if $\A$ has a non-trivial twist symmetry $\Omega$, then the
action is invariant, {\it i.e.} $S(\Psi)=S(\Omega \Psi)$. In this case
it is possible to truncate the string field theory to the $\Omega$
eigenstates of eigenvalue one, thereby preserving the master equation.

In this case we can extend the twist operator to the tensor
theory by defining $\bOmega = \Omega \otimes \omega$, where $\omega$
is a suitably defined twist operator on $\a$. If $\a$ is a direct
sum of matrix algebras, we classify all inequivalent choices for
$\omega$, where we use the fact that a full matrix algebra has no
(linear) outer automorphisms. In the case where $\a$ is a single
matrix algebra, there always exists the twist operator
$\omega_I(A)=A^t$, where $t$ denotes transposition. If the rank of the
matrix algebra is even, there exists a second inequivalent choice
$\omega_J(A)=J_0 A^t J_0$, where $J_0$ is the matrix which is used to
define symplectic 
transposition (see section~5.5). The corresponding twist operators
$\bOmega_I$ and $\bOmega_J$ commute, and we can truncate the theory to
the subspace, where either one or both twist operators have eigenvalue
one. In this way we can obtain open string theories whose gauge groups
are $SO(n)$ (using $\bOmega_I$), $USp(2n)$ (using $\bOmega_J$), 
or $U(n)$ (using both twist operators). Somewhat surprisingly, we 
find that the unoriented $SO(2n)$ and $USp(2n)$ theories 
possess a  nontrivial twist symmetry. 

If the algebra $\a$ is a direct sum of matrix algebras, $\a$
may have in addition outer automorphisms. In this case, there exist
additional inequivalent choices for $\omega$, but they do not 
lead to new constructions.
\smallskip

The above analysis is appropriate for the case where
$\Omega^2=+1$. This is certainly the case for bosonic theories, but it
need not hold for theories containing fermions, as was pointed out by
Gimon and Polchinski \cite{GimPol}. In this case case, $\Omega^2$ only
has to be an automorphism of the algebra $\A$, and this allows for
additional constructions to be discussed in \cite{GZ2}. Similarly, the
original string field theory may have additional symmetries (beyond
the twist symmetry), and other restrictions may be possible. This will
also be discussed in detail in \cite{GZ2}.

\section{The algebraic framework of open strings}
\setcounter{equation}{0}

In this section we shall exhibit in detail the basic algebraic
structure of an open string theory, and we shall explain how to
construct from it a Grassmann even master action satisfying
(i--iii).  Some of the algebraic ingredients were clearly spelled out in
the original construction of Witten \cite{Witten} and elaborated very
explicitly by Thorn in Ref.~\cite{Thorn}, where BV quantization played
an important role. 

The present exposition modifies some aspects of the above, and
incorporates new elements.  For example, it will be recognized that
the ``trace'' is not fundamental, but that it is sufficient to have a
suitable bilinear form. We will also show how to construct a star
operation from the familiar BPZ and hermitian conjugations. Most
importantly, we shall not assume that the underlying algebra is
strictly associative, but rather that it has the structure of a
homotopy associative $A_\infty$ algebra of Stasheff \cite{Stash}. We
shall also reexamine the cyclicity and conjugation properties in this
context.

At the end of the section we shall analyze the conditions that
guarantee that two string field theories are strictly equivalent. 
In particular, we shall show that this is the case if the 
homotopy associative  algebras are homotopy equivalent, where the
homotopy equivalence induces an isomorphism of the underlying vector
spaces and satisfies a cyclicity property.

\subsection{The vector space of the string field}

The vector space that is relevant for the open string algebra is the 
vector space $\A$ of the states of the complete conformal theory
which includes both the matter and the ghost sector. This is a complex
vector space which carries a natural $\Zop$ grading induced by the
ghost sector of the theory\footnote{In principle, the degree has to be
defined separately for the various string theories; it is  typically
convenient to define it in such 
a way that the classical part of the string field is at degree
zero. In bosonic open strings this is achieved by setting 
$\hbox{deg}= 1- \hbox {gh}$, since the classical open string fields
appear at ghost number one. In bosonic string theory, since the ghost
fields are anticommuting, the degree of the various states is
correlated with the even or odd character of the  corresponding vertex
operators. This is not the case for superstrings where the conformal
theory also contains commuting ghosts and anticommuting matter fields. In
this case  the degree still involves the ghost number, but may have   
a contribution from the superghost picture number.}.  
Given a basis $\ket{\Phi_i}$ of vectors of definite grade in $\A$, the
dual basis $\bra{\Phi^j}$ (which is defined by 
$\bra{\Phi^i}\Phi_j\rangle = \delta^i_j$) defines a basis for 
$\A^*$. We define a $\Zop$ grading on $\A^*$ by 
$\epsilon(\bra{\Phi^i}) =- \epsilon (\ket{\Phi_i})$.
When we wish to emphasize the complex field we will write the above
vector spaces as $\A_\Cop$ and $\A^*_\Cop$. 
\medskip

The entire structure of the open string field theory is encoded in
this complex vector space. For certain purposes, however, it is
convenient to extend this vector space to a {\it module} over a
certain Grassmann algebra $\G$; this can be done as follows. First of
all, we introduce a spacetime field $\phi^i$ for every basis vector
$\ket{\Phi_i}$ of $\A_\Cop$ of definite grade. The $\Zop$ grading of
$\A_\Cop$ induces (up to an overall constant) a $\Zop$ grading on the
space of fields, and we can choose this grading in such a way that the
fields corresponding to physical bosons and fermions have even and odd 
degree, respectively. We then define $\G$ to be the Grassmann algebra
which is generated by these fields, where the Grassmannality of $\phi$
is even (odd) if the degree of $\phi$ is even (odd). Finally, we
define $\A_\G$ to be the module over the algebra $\G$ which consists
of the states 
\be
\sum_{i} \ket{\Phi_i} g^i \,,
\ee
where $g^i$ are arbitrary elements in $\G$.\footnote{If $\G$ was a
field rather than only an algebra, $\A_\G$ would be a vector space
over $\G$, whose basis over $\G$ defines a basis of $\A_\Cop$
over $\Cop$.} $\A_\G$ has a natural $\Zop$ degree, where
the degree of $Eg\in\A_\G$ is the sum of the degrees of $E\in\A_\Cop$
and $g\in\G$, 
\be  
\label{grading}
\epsilon (Eg ) \equiv  \epsilon (E) + \epsilon  (g)\,.
\ee
The module $\A_\G$ contains the subspace (over $\Cop$) of {\it string
fields}, which are the states of the form 
\be
\label{stringf}
\sum_{i} \ket{\Phi_i}\, \phi^i \,.
\ee 
We can also introduce the dual module $\A^*_\G$, which is the space
of states
\be
\sum_{i} h_i \bra{\Phi^i} \,,
\ee
where $h_i\in\G$, and the grading is defined by the sum of gradings. 
For $Eg\in\A_\G$ and $hF\in\A^*_\G$ we define the pairing 
$\langle \cdot | \cdot \rangle$ by  
\be
\langle hF | Eg \rangle = h \langle F | E \rangle g\,.
\ee
This defines an element in $\G$, and it satisfies
\be
\epsilon (\langle A | B \rangle) = \epsilon(A) + \epsilon(B) \,,
\ee
where $A\in\A^*_\G$ and $B\in\A_\G$.

\subsection{Symplectic form and BRST operator}

In order to be able to formulate a string action for open string
theory, we need a symplectic bilinear form on $\A_\G$. 
This form is induced from a natural inner product on $\A_\Cop$ which
we examine first.

The bilinear form $\langle \cdot\, , \, \cdot \rangle$ on $\A_\Cop$ 
is a map from $\A_\Cop\otimes \A_\Cop$ to the complex numbers. It
is said to be of degree minus one which means that it is only
nontrivial on pairs of vectors whose degrees add to minus one
\be
\label{minusone}
\langle E_1 , E_2 \rangle  \not= 0  \quad\Rightarrow\quad 
\epsilon(E_1) + \epsilon(E_2) =-1\,,
\ee 
where $E_1,E_2 \in\A_\Cop$. \footnote{By an overall redefinition of
degrees one can change the minus one to any odd number. We therefore
say that the bilinear form is odd.} The bilinear form is defined as an
element $\bra{\omega_{12}} \in\A^*_\Cop \otimes \A^*_\Cop$, where 
the subscripts $1$ and $2$ label the two spaces in the tensor product.
We set  
\be
\label{bpzinm}
\langle E_1 , E_2\rangle \equiv  \bra{\omega_{12}} E_1\rangle_1
\ket{E_2}_2\,. 
\ee
Bilinearity demands that for $a,b\in\Cop$
\be
\label{billinear}
\langle E a , F b \rangle = \langle E , F\rangle  \,a \, b \,.
\ee
Since $\langle ., .\rangle$ only couples vectors of opposite $\Zop_2$
grade, we say that the $\Zop_2$ grade of $\bra{\omega_{12}}$ is odd. 
In addition, we require that
$\bra{\omega_{12}} = -\bra{\omega_{21}} $, 
which together with (\ref{bpzinm}) implies that 
\be
\label{exchgp}
\langle E , F\rangle 
= -  (-)^{\epsilon(E) \epsilon(F)} \langle F , E\rangle 
= -  (-)^{EF}\langle F , E\rangle \,,
\ee
where we use the convention that $E=\epsilon(E)$ in the exponent.
The sign factor here is a bit redundant since 
$\epsilon(E)\epsilon(F)$ is odd whenever the
form is nonvanishing, but it will be useful for later purposes to keep
it. Finally, the bilinear form is invertible; there is an 
$\ket{s_{12}} \in \A_\Cop\otimes \A_\Cop$ satisfying
$\bra{\omega_{12}}s_{23}\rangle = {}_3{\bf 1}_1 $,
and $\ket{s_{12}}= \ket{s_{21}}$. The $\Zop_2$ 
grade of $\ket{s_{12}}$ is also odd.
\medskip

We extend this bilinear form to $\A_\G$ by defining
\be
\label{bilinear}
\langle E\lambda , F\mu\rangle = \langle E , F\rangle  \,\lambda \, 
\mu \, (-)^{\lambda F}\,,  
\ee
where $E,F\in\A_\Cop$ and $\lambda,\mu\in\G$. 
This defines a bilinear map
$\A_\G \o \A_\G \rightarrow \G$, which we can formally write as
\be
\label{bpzinn}
\langle A, B \rangle = \langle \omega_{12} \ket{A}_1 \ket{B}_2 \,.
\ee
On $\A_\G$, the bilinear form does not necessarily vanish if the
grades of the vectors do not add up to $-1$; instead, we have 
\be
\label{oddf}
\epsilon\, (\,\langle A , B\rangle\,) = \epsilon (A) + \epsilon (B) +1\,,
\ee
where the grading on the left hand side is the grading in $\G$. 
The exchange property now takes the form
\be
\label{exchgpr}
\langle A , B\rangle = -  (-)^{AB}\langle B , A\rangle \, .
\ee
The form is ``symplectic" in that it is antisymmetric when the vectors
are even. 

{}From now on we shall be referring to $\A_\G$, whenever we write
$\A$. For simplicity, we shall be formulating the various properties
directly in terms of $\A_\G$, rather than giving them on $\A_\Cop$,
and explaining how they can be extended to $\A_\G$. 
\medskip

There exists an operator $Q$ of degree $-1$ which squares to zero,
$Q^2=0$. This operator is called the BRST operator, and satisfies
$\bra{\omega_{12}} (Q_1 + Q_2) =0$,
which implies that
\be
\label{qinnp}
\langle QA , B \rangle = -(-)^A \langle A , QB \rangle\,.
\ee
Combining this result with (\ref{exchgpr}) we conclude that
\be
\label{twpone}
\langle A , QB \rangle = (-)^{AB + A+B} \, \langle B, QA \rangle \,,
\ee
and this equation can be thought of as governing
the cyclicity property  of the bilinear form 
$\langle  \cdot \,, Q \, \cdot \rangle$. 
The invertibility of the symplectic form implies that
$(Q_1 + Q_2)\, \ket{s_{12}} =0$.
 
It is possible at this early stage to conclude that the string field 
must be an even element of $\A_\G$.
This is required in order to have a nonvanishing kinetic term. 
Indeed, the kinetic term is of the form $\langle \Phi , Q \Phi\rangle$,
and it therefore vanishes unless $\Phi$ is even since
\be
\label{sfeven}
\langle \Phi , Q \Phi\rangle = - \langle Q \Phi , \Phi\rangle 
= (-)^\Phi \langle \Phi , Q \Phi\rangle\,,
\ee
where use was made of the exchange property (\ref{exchgpr}) and of
the BRST property (\ref{qinnp}).

Some of the axioms can be changed by ``suspension'', {\it i.e.} by
redefining the grading in $\A_\Cop$, and therefore consequently in
$\A_\G$, by an arbitrary integer. 
The grading in $\G$, however, can only be changed by an even
integer, as the $\Zop_2$ grade is determined by the fermionic or
bosonic character of the field. For example, we can change the grading
in $\A_\Cop$ so that the string field becomes odd, and the symplectic
form satisfies  $\bra{\omega_{12}} = + \bra{\omega_{21}}$, among
others. (At a concrete level this change could be implemented by
redefining the degree of the vacuum state.) However, we cannot
change the property that the symplectic form is odd, since after
suspension, it still couples only even vectors to odd vectors in
$\A_\Cop$.

\subsection {BPZ conjugation}

BPZ conjugation is an odd linear map from ${\cal A}$ to $\A^*$,
$\ket{A}\mapsto \bra{\hbox{bpz} (A)}$, and it is defined by
\be
\label{innpr}
\bra{\hbox{bpz} (A)} B\rangle = \langle A, B\rangle \,.  
\ee 
In view of (\ref{bpzinn}) this is equivalent to 
${}_2\bra{\hbox{bpz} (A)} \equiv  \bra{\omega_{12}} A\rangle_1$.
As $\epsilon (\langle B | C \rangle) = \epsilon(B) + \epsilon(C)$ it 
follows that BPZ has degree one, {\it i.e.}
$\epsilon(\bra{\hbox{bpz}(A)})=\epsilon(\ket{A})+1$.

BPZ conjugation has an inverse, an odd linear map bpz${}^{-1}$ from
$\A^*$ to $\A$, denoted as $\bra{A} \mapsto \ket{\hbox{bpz}^{-1}(A)}$
and defined by
\be
\label{bpzinv}
\ket{\hbox{bpz}^{-1}(A) }  \equiv  {}_2\bra{A} s_{12}\rangle \,
(-)^{A+1}\,, 
\ee
where the sign factor ($A$ is the degree of $\bra{A}\in\A^*$) 
is required to insure that bpz $\circ$ bpz${}^{-1}=$
bpz${}^{-1}\circ$ bpz$=\bbbone$, as the reader may verify.  
Having an inverse map, equation (\ref{innpr}) can also be written as  
\be
\label{innprd}
\bra{A} B\rangle = \langle \hbox{bpz} ^{-1}(A), B\rangle \, .
\ee
It also follows from the above definitions that
\be
\label{tensgr}
\begin{array}{lcl}
\hbox{bpz} ( \ket{A} \lambda ) &=& (\hbox{bpz}(\ket{A}))\, \lambda
\\ \hbox{bpz}^{-1} ( \lambda \bra{A} ) &=& (-)^{\lambda} \, \lambda\,
\hbox{bpz} ^{-1}( \bra{A}) \,,
\end{array} 
\ee
where $\lambda\in\G$.

\subsection{Hermitian conjugation}

Next we want to assume that there exists a hermitian conjugation
map ``hc'' from $\A$ to $\A^*$ which satisfies  
\be
\label{hccon}
\overline{\bra{\hbox{hc}(A)}B\rangle} = \bra{\hbox{hc}(B)} A \rangle \,, 
\ee 
where the overline denotes complex conjugation in $\G$. (Recall that
for a Grassmann algebra complex conjugation satisfies  
$\overline {\chi_1 \chi_2} = \bar\chi_2 \bar\chi_1$). hc is an
antilinear map from $\A$ to $\A^*$ taking
\be
\label{hccoon}
\hbox{hc}: \, \ket{A} \lambda  \mapsto \bar\lambda \,
\bra{\hbox{hc}(A)} \,.
\ee
We shall assume that hc is of degree one, {\it i.e.} that
$\epsilon (\bra{\hbox{hc}(A)} ) = \epsilon(\ket{A})+1$. Furthermore,
hc is invertible with inverse hc${}^{-1}:\A^* \to \A$ 
\be
\label{hcmcon}
\hbox{hc}^{-1}: \, \lambda\bra{A} \mapsto  \, \bra{\hbox{hc}^{-1}(A)}
\bar\lambda\,.
\ee
Finally, the BRST operator is required to be hermitian, in the sense
that  
\be
\label{hermq}
\hbox{hc} ( Q \ket{A} ) =  \bra{\hbox{hc}( A) }  Q \, .
\ee

\subsection{Star conjugation}

We now have two (degree one) maps taking $\A$ to $\A^*$, bpz, and hc,
and two (degree minus one) maps taking $\A^*$ to $\A$, namely
bpz${}^{-1}$, and hc${}^{-1}$.  We can use these maps to construct a
star conjugation $*$, an antilinear, even map from $\A$ to $\A$. There
are two obvious candidates 
\be
\label{twomaps}
\hbox{bpz}^{-1} \circ \hbox{hc} \quad  \hbox{and}, \quad
\hbox{hc}^{-1}\circ \hbox{bpz} \,.
\ee
These two maps are inverses of each other, and they
are relevant in the analysis of the reality of the symplectic
form. Indeed, it follows from  eqns.~(\ref{innpr}) (\ref{hccon}) and
(\ref{innprd}) that 
\begin{eqnarray}
\label{howdef}
\overline {  \langle A , B \rangle } & = &  \overline{ 
\bra{\hbox{bpz}(A)} B \rangle } \nonumber \\ 
&=&  \bra{ \,\hbox{hc} (B)} \,\hbox{hc}^{-1} \circ \hbox{bpz} (A)
\rangle \\ 
&=&  \langle \,\hbox{bpz}^{-1} \circ \hbox{hc} (B) \,,\, 
 \hbox{hc}^{-1} \circ \hbox{bpz} (A) \,\rangle \,. \nonumber
\end{eqnarray}
It is clear that if we require the two maps in (\ref{twomaps})
to be identical we obtain a single star conjugation that squares to
one, as it should, and controls the reality property of the
symplectic form. We therefore demand 
\be
\label{twomapseq}
* = \, \hbox{bpz}^{-1} \circ \hbox{hc} \,=\, \hbox{hc}^{-1}\circ
\hbox{bpz}\,. 
\ee
This star operation satisfies $\epsilon (A^*) =\epsilon(A)$ and
\be
\label{invol}
(A^*)^* = A\, .
\ee
Using (\ref{howdef}), the equality of the maps  now implies
\be
\label{hermsym}
\overline {  \langle A , B \rangle }  = \langle B^* , A^* \rangle \,.
\ee
This is the  conjugation property of the  bilinear form. 
It is a simple exercise to show that
\be
\label{conjgr}
(A\lambda)^* = (-)^{\lambda (A+1)} \, A^* \bar \lambda  \,.
\ee
The hermiticity of the BRST operator (\ref{hermq}) now implies that
\begin{eqnarray}
(Q \ket{A} )^* &=&  \hbox{bpz}^{-1} \circ \hbox{hc} ( Q \ket{A})  ) =  
\hbox{bpz}^{-1}  ( \bra{\hbox{hc} (A)}  Q  )  \nonumber \\
&=&  (-)^{A+1} {}_2\bra{\hbox{hc}(A)} Q_2 \ket{s_{12}} = 
-Q_1 \,\, {}_2\bra{\hbox{hc}(A)}  s_{12}\rangle \nonumber \\
&=& -(-)^A  Q \bigl (  (-)^A  {}_2\bra{\hbox{hc}(A)}  s_{12}\rangle
\bigr) =  -(-)^A  Q \ket{A^*}\,, \nonumber
\end{eqnarray}
and therefore 
\be
\label{qstar}
(QA)^* = -(-)^A \, Q A^*\,.
\ee
It follows from (\ref{hermsym}) and (\ref{qstar}) that
\be
\label{kinstar}
\overline { \langle A , QB \rangle } = \langle B^* , Q A^*\rangle\,.
\ee

\subsection{Reality of the string field kinetic term}

As explained in subsection~2.1, the string field defines a certain
subspace of $\A_\G$. Having constructed a star conjugation on $\A_\G$,
we can impose a reality condition on the string field. This will
take the form $\Phi^* = \pm \Phi $, where the top sign 
corresponds to a ``real" string field and the lower sign to an
``imaginary" string field. Either condition guarantees the
reality of the kinetic term, as (\ref{kinstar}) implies 
\be
\label{realkin}
\overline{  \langle \Phi , Q \Phi \rangle } =  \langle \Phi ^*, Q \Phi
^*  \rangle 
= \langle \Phi , Q \Phi \rangle \, .
\ee
For definiteness we shall take the string field to be real  
\be
\Phi^* = \Phi \, .
\ee
Using a basis $E_i$ of real vectors ($E_i^* = E_i$) for $\A_\Cop$,
the string field can be expanded  as 
\be
\label{rsf}
\Phi = \sum_i  E_i \, \phi^i  \,,
\ee
where $\phi^i\in\G$. As the string field is even, the degree
of $E_i$ is the same (mod $2$) as that of $\phi^i$; because of 
(\ref{conjgr}), the reality condition on $\Phi$ then requires that
$\overline {\phi^i} = \phi^i$.

\subsection{$A_\infty$ algebras}   

In order to have an interacting theory we need to introduce an 
algebra structure on $\A_\G$. We first  explain how the 
$\Zop$ graded (complex) vector space $V=\A_\Cop$ can be given
the structure of a homotopy associative algebra, and then
extend the construction to 
$\A_\G$. A homotopy associative algebra is a natural generalization
of an associative differential graded algebra in which the condition of
associativity is relaxed in a particular way. Most of what we do in
this subsection is to review the definition of an $A_\infty$
algebra. We follow the discussion of 
Ref.~\cite{GJ}. (Strictly speaking, we will be describing the
bar construction associated to a homotopy associative algebra; the
conventional description is given in appendix~A.)

Given a $\Zop$ graded  complex vector space $V$, we consider the 
canonical  tensor co-algebra 
$T(V) = V\oplus (V\otimes V) \oplus (V\otimes V\otimes V)\oplus\cdots$ 
with $\Zop$ gradings obtained from those on $V$. Let us assume that
we are given a collection of multilinear maps $b_n: V^{\o n}\to V$,
where $n=1,2, \ldots$, all of which are of degree $-1$. We use these
maps to define a linear map  ${\bf b} : T(V) \to T(V)$ of degree $-1$;
acting on the $V^{\o n}$ subspace of $T(V)$ it is defined as    
\be
\label{puttog}
{\bf  b} =\sum_{i=1}^n  \, \sum_{j=0}^{n-i}  
\bbbone^{\otimes j} \otimes b_i \otimes \bbbone^{n-i-j}\, , \quad
\hbox{on} \,\, V^{\otimes n}\,.  
\ee
By virtue of this definition, ${\bf b}$ is a coderivation
\cite{GJ}. As ${\bf b}$ is of odd degree, its square ${\bf b}^2$ is
also a coderivation. We now demand that ${\bf b}^2 =0$, {\it i.e.} that 
${\bf b}$ is a differential. Since ${\bf b^2}$ is a coderivation, it
is sufficient to demand that the multilinear maps associated to 
${\bf b^2}$ vanish, {\it i.e.}
\be
\label{onvn}
\sum_{i =1}^{n} b_i \circ b_{n+1-i} = 0 \, \quad \hbox{on} \,\,
V^{\otimes n} \,. 
\ee
Here we have used the notation that $b_k$ acts on $V^{\o n}$ for $n>k$
as the sum of all suitable 
$\bbbone^{\otimes p} \otimes b_k \otimes\bbbone^{\otimes q}$, where 
$p+k+q=n$. This is best illustrated by writing the first few
identities of (\ref{onvn}) explicitly
\be
\label{tryit}
\begin{array}{lllcl}
\hbox{On} \, V & : \quad & 0 &=& b_1 \circ b_1 \\
\hbox{On}  \, V^{\otimes 2}& : &  0 &=& 
b_1 \circ b_2 + b_2 \circ 
(b_1 \otimes \bbbone + \bbbone \otimes b_1 ) \,,  \\
\hbox{On}  \, V^{\otimes 3} & : &  0 &=& 
b_1 \circ b_3 + b_2 \circ 
(b_2 \otimes \bbbone + \bbbone \otimes b_2 ) \,,  \\
& & & &  + b_3 \circ ( b_1 \otimes \bbbone\otimes \bbbone
+ \bbbone \otimes b_1 \otimes \bbbone +  \bbbone \otimes
\bbbone\otimes b_1) \,.
\end{array}
\ee
The map $b_1$ is just the BRST operator $Q$, and we write 
$b_1(A) =QA$. $b_2$ is the product in the algebra, which we
write as $b_2 (A,B) = AB$, and $b_3$ is the first of the
``higher homotopies", which is denoted by $b_3(A,B,C)= (A,B,C)$. 
The explicit form of the above identities acting on vectors is readily
found. There are obvious sign factors that arise as the $b$'s (which
are odd) are moved across the vectors; in more detail (\ref{tryit})
becomes then 
\begin{eqnarray}
\label{tryy}
0 &=&  Q^2 A\nonumber \\
0 &=& Q (AB) +  (QA)B+ (-)^A A (QB) \,,  \nonumber \\
0 &=& Q (A,B,C) + (AB)C + (-)^A A(BC) \,,  \nonumber \\
& {} &  + (QA,B,C)   + (-)^A (A,QB,C)  
+ (-)^{A+B} (A,B,QC)   \,.
\end{eqnarray}
The second identity essentially expresses the fact that $Q$
acts (up to signs) as an odd derivation on the product.  The last
equation indicates that the algebra fails to be associative up to
terms involving the homotopy $b_3$ and the BRST operator.

The above construction is easily extended to $\A_\G$ 
by defining 
\begin{eqnarray}
b_n (A_1 , \cdots , (A_i \lambda_i + B_i \mu_i) ,  \cdots , A_n ) 
&\equiv  &
{\displaystyle (-)^{\lambda_i  (A_{i+1} +\cdots A_n)} 
b_n (A_1 , \cdots , A_i ,  \cdots , A_n )\,\lambda_i} \nonumber \\
& &  {\displaystyle + (-)^{\mu_i  (A_{i+1} +\cdots A_n)} 
b_n (A_1 , \cdots , B_i ,  \cdots , A_n )\,\mu_i \,, } \nonumber
\end{eqnarray}
where $\lambda_i, \mu_i \in \G$. It is clear that the extended
products still obey the main equations (\ref{onvn}).

\subsection{Cyclicity and star conjugation}  

Now that we have introduced on $\A_\G$ the (homotopy) algebra
structure, necessary to give interactions to the open string theory,
we must examine  what extra conditions must be imposed  on the algebra
in order to be able to write a consistent string action. 
As will become clear from the analysis in 
subsection~2.9, 
we shall have to require that the products together with the symplectic form
have suitable graded cyclicity properties in order to be able to write
an action that satisfies the master equation.\footnote{In the strictly
associative case this is familiar \cite{Witten,Thorn}. For closed strings
the analogous requirement was explained in Ref.\cite{Z1} Sect.4.2,
and elaborated mathematically in \cite{alexandroff}. } Furthermore,
the star  conjugation must have a well defined action on the products if
we are to be able to guarantee that the interacting action is
real. In this subsection we shall explain the two conditions in turn;
these will be used in the following subsection to show that a
consistent string action can be obtained.

As regards cyclicity the first case was already discussed in 
deriving (\ref{twpone}) which can be rewritten as 
\be
\label{fstcy}
\langle A_1 , b_1 (A_2) \rangle = (-)^{A_1 + A_2 + A_1A_2 } 
\langle A_2 , b_1 (A_1) \rangle \,.
\ee
We shall now demand that this property extends to the product $b_2$
and the higher homotopies as
\be
\label{cyclmn}
\langle A_1 , b _{n}(A_2 , \cdots , A_{n+1} ) \rangle = (-)^{s_{n+1}(A)}
\langle A_2, b _{n} \, (A_3, \cdots , A_{n+1}, A_1  )\rangle \,,
\ee
where the sign factor is 
\be
\label{sgnfh}
s_{n+1}( A_1 ,  A_2 , \ldots\,  ,A_{n+1} ) =A_1 + A_2 + A_1 (A_2 + A_3  +
\cdots + A_{n+1})  \,. 
\ee
The sign factor can be identified as the sign necessary to rearrange
the vectors into the final configuration, taking into account
that $b_n$ is odd. For $n=1$, (\ref{cyclmn}) reduces to
(\ref{fstcy}), and after $n+1$ applications, the expression comes back
to itself. We should mention that (\ref{sgnfh}) is uniquely determined
for $n\geq 2$ by the choice for $n=1$ (which corresponds to
(\ref{fstcy})) and the identities of the homotopy  associative
algebra. Indeed, we can use (\ref{cyclmn}) (together with the exchange
property of the symplectic form if necessary) to  rewrite each term in
the identity  
\be
\label{checkc}
\Bigl\langle  A_1 \,,  \sum_{l=1}^{n-1} b_l \circ b_{n-l} \, (A_2 ,
\ldots , A_n ) \Bigr\rangle = 0 \,, 
\ee
in terms of expressions of the form 
\be
\langle A_2 , b_{p} (A_3, \ldots, A_r, 
b_{n-p}(A_{r+1}, \ldots , A_{r+n-p}), A_{r+n-p+1}, \ldots, A_n, A_1)
\rangle\,.
\ee
These terms have to reassemble again (up to an overall sign) into an
identity of the form (\ref{checkc}), and this requirement determines
the sign (\ref{sgnfh}) uniquely.
\smallskip

The situation is somewhat similar in the case of the
star conjugation. As we have discussed before in (\ref{qstar}), the
star conjugation  acts on $b_1 =Q$ as
\be
\label{qstarp}
(b_1 (A))^* = (-)^{A+1} b_1 ( A^*) \,.
\ee
We want to extend this definition to the higher products, where we 
demand that it acts on products of vectors (up to a sign factor)
by reversing the order and taking the star conjugation of each 
vector separately. 
This definition must be compatible with the
identities of the homotopy associative algebra, and we find that 
this consistency condition determines the sign factor
in the star conjugation property as
\be
\label{homstar}
\bigl( b_n( A_1, \ldots, A_n )  \bigr) ^* = (-)^{A_1 + \cdots A_n +1}
b_n ( A_n^* , \ldots , A_1^*)\,. 
\ee
The sign factor is uniquely determined up to the trivial ambiguity
which corresponds to the redefinition of $*$ by $-*$. This would 
not change the BRST property, nor the fact that star squares to the 
identity operator, but it would change the sign factor in
(\ref{homstar}) by $-1$ for even $n$. This change can be 
absorbed into a redefinition of $b_n$ by $b_n \to (i)^{n+1} b_n$ which
does not change the identities of the homotopy associative algebra.
A more structural analysis of the sign factor can be found in
appendix~B.

\subsection{Master action and reality}

In the present section we will write the classical master action,
and show that it satisfies the classical master equation of
Batalin and Vilkovisky by virtue of the homotopy algebra identities
and the cyclicity properties of the bilinear form.
We will see that the reality condition for the string field
guarantees that the master action is real, and we will explain why
it is possible to impose a reality condition preserving the fact that
the action satisfies the master equation.  

We claim that the classical master action $S(\Phi)$ is given by
\begin{eqnarray}
\label{maction}
S &=&  \sum_{n=1}^\infty {1\over n+1}
\langle \Phi \,, b_n (\Phi,\,\ldots \,,\Phi) \rangle\nonumber \\
 &=&  {1\over 2} \langle \Phi \,,  Q \Phi\rangle
  + {1\over 3} \langle \Phi, \Phi \Phi\rangle
+ {1\over 4} \langle \Phi ,  (\Phi, \Phi, \Phi)\rangle  + \cdots  
\end{eqnarray}
We must now verify that $S$ satisfies the master equation
\be
\label{masteqn}
\{ S , S \} \equiv {\partial_r S \over \partial \phi^i} \, \omega^{ij}
{\partial_l S \over \partial \phi^j}  = 0\,,
\ee
where 
$\partial_r$ and $\partial_l$ denote the right and left derivative,
respectively, and we have set 
\be
\label{notation}
\Phi = E_i \phi^i \,, \quad  \omega_{ij} = \langle E_i , E_j \rangle\,,
\ee
and $E_i$ is our real basis in $\A_\Cop$. As a first step in verifying
(\ref{masteqn}) we calculate the variation of the action under a
variation of the string field $\Phi$. Using the cyclicity properties
of the bilinear form, we find
\be
\label{vary}
\delta S = \langle \delta \Phi , V(\Phi) \rangle\,, 
\ee
where
\be
\label{emotion}
V\,(\Phi) = \sum_{n=1}^\infty b_n (\Phi , \ldots  , \Phi)
= Q\Phi + \Phi \Phi  + (\Phi , \Phi, \Phi) + \cdots \,.
\ee
If we use (\ref{vary}) and (\ref{notation}) to write
\be
\label{prepa}
\delta S = \delta \phi^i\,  \langle  E_i\,  , V(\Phi) \rangle 
= -\langle V(\Phi)\, ,\, E_i \rangle\,  \delta \phi^i \,, 
\ee
it is then simple to see that
\be
\label{deriv}
{\partial_r S \over \partial \phi^i} = -\, \langle V(\Phi)\, ,\, E_i\,
\rangle\,,  
\qquad
{\partial_l S \over \partial \phi^j} = \,\langle\,E_j\,  , V(\Phi)
\rangle \,, 
\ee
and the master equation becomes 
\be
\label{meqn}
\{ S , S \} =  -\,\langle V(\Phi)\, ,\, E_i\, \rangle \,\omega^{ij}
\,\langle\,E_j\,  , V(\Phi) \rangle = -\langle V(\Phi), V (\Phi)
\rangle\,, 
\ee
where in the last step  we have recognized that  
the component  $V^i$ of $V$ defined by the expansion
$V=E_i V^i$,  is given as  
$V^i = \omega^{ij} \langle E_i , V \rangle$. 
It follows that it is sufficient to show that
\be
\sum_{n, m =1}^\infty \langle b_n (\Phi, \ldots ,\Phi) \,,\, b_m
(\Phi, \cdots ,\Phi) \, \rangle = 0 \,. 
\ee
The sums in this equation can be rearranged to read
\be
\label{testeq}
\sum_{n=1}^\infty \sum_{l=1}^{n} \langle \,b_l  \,,\,
b_{n+1-l}\,\rangle  = 0 \,,
\ee
where we have omitted the string field $\Phi$ for simplicity.
We shall now show that (\ref{testeq}) holds, as each term (in the sum
over $n$) vanishes for fixed $n$. This is manifestly true for $n=1$,
as $\langle  b_1 (\Phi) , b_1 (\Phi)\rangle = 
- \langle \Phi , b_1 b_1 (\Phi) \rangle =0$. 
In general, we claim that  
\be
\sum_{l=1}^{n} \langle \,b_l  \,,\, b_{n+1-l} \, \rangle
= - {2\over n+1}\, \sum_{l=1}^{n}\,\langle\, \cdot \,, b_l\circ
b_{n+1-l}\rangle =0 \,,
\ee
where we have used, in the last equation, the $b$ identities of the
homotopy algebra. The first equation follows from the fact that
the terms 
$\langle b_l,b_{n+1-l}\rangle$ and $\langle b_{n+1-l}, b_l \rangle$
are actually identical, and that their sum can be rearranged into the
form 
$\langle\,\cdot\,,\,(b_l\circ b_{n+1-l}+b_{n+1-l}\circ b_l)\rangle$. 
There are $(n+1)$ terms of this form, and this leads to the claimed
proportionality factor.  
\medskip

We observe that the master equation (\ref{masteqn}) holds in general
for complex fields $\phi^i$ 
as the reality condition has not been used so far. Furthermore,
it follows from the explicit form of the master action that it is an
analytic function of the complex field $\phi^i$. In particular,
this implies
that the analytic function $\{S , S\}
=(\partial S/\partial \phi^i )\omega^{ij} 
(\partial S/\partial \phi^j)$ vanishes identically.\footnote{ 
The $\omega^{ij}$'s are analytic functions of the fields, in fact, 
in our case they are just constants.}
It is 
then clear that $\{S, S\}$ also vanishes if the fields
$\phi^i$ are restricted to be real. In particular, this 
means that the action, restricted to real fields, still
satisfies the master equation, and therefore that it is possible to
impose consistently the reality condition on the string field.

Once we impose the reality condition $\Phi^* = \Phi$ on the
string field, the master action (\ref{maction}) is
real because each term is real. Indeed
\begin{eqnarray}
\label{echtrm}
\overline {\langle \Phi , b_n (\Phi, \ldots , \Phi ) \rangle}
&=& \langle  (b_n (\Phi, \ldots , \Phi ))^* , \Phi^*  \rangle \,,
\nonumber \\ 
&=& -\langle  b_n (\Phi^*, \ldots , \Phi^* ) , \Phi^*  \rangle\,,
\nonumber \\  
&=& \langle  \Phi^* , b_n (\Phi^*, \ldots , \Phi^* )  \rangle \,
,\nonumber \\ 
&=& \langle  \Phi , b_n (\Phi, \ldots , \Phi )  \rangle \,,
\end{eqnarray}
where we have used the conjugation property (\ref{homstar}) and the
exchange property of the bilinear form. 
As we have chosen the string field to be real and the action as given 
above, the reality of  the action fixes now the ambiguity in the
definition of the star operation for the string products. Conversely,
had we chosen a different convention for the star operation, we
would have had  to introduce suitable factors of $i$'s into the
interaction terms to guarantee reality; these factors could then be
absorbed into a redefinition of the  $b_n$'s,  as explained before.

\subsection{Equivalent field theories and homotopy equivalence 
of $A_\infty$ algebras}

In this subsection we shall show that two string field theories are
equivalent if the corresponding homotopy associative are homotopy
equivalent, where the homotopy equivalence induces an isomorphism of
the underlying vector spaces, and satisfies a certain cyclicity
condition. 
We shall not assume that the two theories are related by an
infinitesimal deformation; our analysis is therefore more general than
previous discussions of equivalence \cite{hatazwiebach,senzwiebach}.

In the covariant description of the Batalin Vilkovisky formalism 
(see \cite{aschwarz,wittenbv,hatazwiebach, alexandroff}), 
a classical field theory is defined by the
triple $(M,\omega, S)$, where $M$ is a (super)manifold of field  
configurations, $\omega$ is an odd symplectic form on $M$ ({\it i.e.}
$\omega$ is closed and nondegenerate), and the action $S$ is a
function on $M$ satisfying the BV master equation $\{ S , S\} =0$. The
action $S$ defines an odd Hamiltonian vector field $V$ via 
$i_V \omega = -dS$, and the Lie-bracket of the vector $V$ with itself
vanishes $\{V , V\} =0$.

Two field theories $(M,\omega, S)$ and 
$(\tilde M,\tilde\omega, \tilde S)$ are said to be equivalent if
there exists a diffeomorphism $f:M\to \tilde M$ such that 
$\omega= f^* \tilde\omega$ and $S= f^* \widetilde S$ \cite{senzwiebach}.  
The first condition means that $f$ is compatible with the symplectic
structures, and the second implies that the two actions are mapped
into each other. It also follows from these relations that the vectors
$V$ and  $\widetilde V$ are related as $\widetilde V = f_* V$.

In the case of an open string field theory the manifold $M$ is the
complex vector space of string fields,  
and $\omega$ is the
odd bilinear form $\langle \cdot , \cdot\rangle$. In this case the
vector field $V$ is defined by (\ref{emotion}) 
\be
\label{vfield}
V(\Phi) = \sum_{n=1}^\infty b_n  (\Phi, \cdots , \Phi) \,,
\ee
as follows from eqn.~(\ref{deriv}).
Suppose now that we have another open string theory with string field
$\widetilde \Phi$, and vector field $\widetilde V$ 
\be
\label{vfieldt}
\widetilde V(\widetilde \Phi) = \sum_{n=1}^\infty
\widetilde b_n (\widetilde\Phi,\cdots \,,\widetilde\Phi) \,,
\ee
and suppose that the two string field theories are equivalent in the
above sense. Then there exists a family of multilinear maps
$f_n : \A^{\o n} \to \widetilde \A$ of degree zero such that
\be
\label{fieldrs}
\widetilde \Phi = \sum_{n=1}^\infty f_n (\Phi, \cdots \,, \Phi) =
 f_1 (\Phi)  + f_2 ( \Phi , \Phi ) + f_3(\Phi , \Phi , \Phi )
+ \cdots \,.
\ee
In order for this to define a diffeomorphism, a necessary condition is
that $f_1$ defines an isomorphism of the string field vector spaces.
If we choose a basis for these vector spaces, the vectors $V$ and
$\widetilde V$ are then related as
\be
 {\partial_r \widetilde\Phi^i\over
\partial \Phi^j} \, V^j (\Phi)  =  \widetilde V^i (\widetilde \Phi) \, . 
\ee  
Using the definition of $V$ in terms of $b_n$, and $\widetilde V$ in
terms of $\widetilde b_n$ together with (\ref{fieldrs}), this becomes
\be
\sum_{n=1}^\infty [ f_n (V,  \Phi , \cdots ,  \Phi ) + 
f_n ( \Phi ,V, \cdots , \Phi) + \cdots
+ f_n ( \Phi , \cdots , V) ]
 = \sum_{n=1}^\infty \widetilde b_n ( \widetilde \Phi , \cdots
\widetilde\Phi) \,.  
\ee
Finally, we can express $\widetilde\Phi$ in terms of $\Phi$ in the
left hand side, and use (\ref{vfield}) in the right hand side to obtain
the series of identities  
\be
\label{gensete}
\sum_{l=1}^n f_l \circ b_{n+1-l} = \sum_{l=1}^n \widetilde b_l
\sum_{i_1+\cdots +i_l=n}
( f_{i_1} \o \cdots\o  f_{i_l} )\,,
\ee
where we have used the by now familiar shorthand notation. Strictly
speaking, the above argument only implies (\ref{gensete}) when
evaluated on $\Phi^{\o n}$. However, in order for the equivalence to
be an equivalence of open string field theories, it has to respect the
structure of the underlying homotopy associative algebras, and we have
to demand that (\ref{gensete}) extends to the whole of $\A^{\o n}$.

We can expand these equations in powers of $\Phi$, and collect terms
of a given order in $\Phi$. For example, to first (and lowest) order  
in $\Phi$ we have 
\be
\label{feqq}
 f_1 b_1 = \widetilde b_1 f_1 \,,
\ee 
which is the statement that $f_1$ is a cochain map, {\it i.e.} that it
commutes with the differentials. This equation implies that $f_1$
induces a well defined map from the cohomology $H(\A)$ to the
cohomology $H(\widetilde\A)$. The equation at second order is
\be
\label{seceqn}
f_1 b_2 + f_2 (b_1 \o \bbbone + \bbbone \o b_1) = \widetilde b_1 f_2 +
\widetilde b_2 
(f_1 \o f_1)\,.
\ee
This equation implies that $f_1$ actually defines a
homomorphism $H(\A)\to H(\widetilde\A)$ of associative algebras. 

A family of functions $f_n$ satisfying (\ref{gensete}) defines a
{\it homotopy map} from the homotopy algebra $\A$ (defined by the
maps $b_n$) to the homotopy algebra $\widetilde\A$ (defined by the
maps $\tilde b_n$). Two homotopy associative algebras $\A$ and
$\widetilde \A$ are {\it homotopy equivalent} \cite{Stasheff}, if
there exists a homotopy map from $\A$ to $\widetilde \A$, such that 
$f_1$ induces an isomorphism of the cohomologies.\footnote{In this case, 
there always exists a homotopy map from $\widetilde \A$ to $\A$
inducing the inverse isomorphism in cohomology \cite{Stasheff}.} 
We have thus shown that the homotopy associative algebras of two
equivalent open string field theories are homotopy equivalent. In
addition, as we have explained above, $f_1$ must be an isomorphism of
the vector spaces $\A$ and $\widetilde A$.  
\medskip

Next we want to analyze the condition that the diffeomorphism $f$  
respects the symplectic
structure $\omega = f^* \widetilde \omega$, {\it i.e.} that
\be
\label{omegacon0}
\omega_{ij} = {\partial_l \widetilde\Phi^p \over \partial \Phi^i}
\, \widetilde\omega_{pq} \,
{\partial_r \widetilde\Phi^q \over \partial \Phi^j}\,.
\ee
Using the expression of $\widetilde\Phi$ in terms of $\Phi$, we find a
series of conditions, the first of which is
\be
\label{invariant}
\langle f_1(A_1), f_1(A_2) \rangle  =  \langle A_1, A_2 \rangle \,,
\ee
where, by abuse of notation, we have denoted the bilinear forms on
$\A$ and $\widetilde A$ by the same symbol. The higher conditions are
\be
\label{omegacon}
\sum_{l=1}^{n-1} \sum_{r=1}^{l} \sum_{s=1}^{n-l}
\langle f_l (\Phi, \ldots,  A, \ldots , \Phi), 
f_{n-l} (\Phi, \ldots,  B, \ldots, \Phi)\rangle =0 \,, \quad n\geq 3\,,
\ee
where $A$ and $B$ are inserted at the $r^{th}$
and  $s^{th}$ position, respectively. This identity holds
provided that
\be
\label{fcyclic}
\sum_{l=1}^{n-1}  \langle f_l (A_1, \ldots, A_l), 
f_{n-l} (A_{l+1}, \ldots, A_{n})\rangle =0\,.
\ee
Indeed, if we sum the identities (\ref{fcyclic}) over all different
combinations of $A_1, \ldots, A_n$, where $A_i=A$, $A_j=B$ for
$i<j$, and $A_k=\Phi$ for $k\ne i,j$, we reproduce the identities
(\ref{omegacon}), apart from terms, where $A$ and $B$ appear both in
$f_l$ or both in $f_{n-l}$. However, these terms occur in pairs
(because of the symmetry $l\leftrightarrow n-l$), and
the sum of the two terms in each pair vanishes, as the bilinear form
is odd under exchange (\ref{exchgpr}), and $f_n$ and $\Phi$ are even.

The condition (\ref{fcyclic}) can be regarded as a generalization of a
cyclicity condition; indeed, apart from the higher correction terms 
({\it i.e.} the terms where $2\leq l \leq n-2$), it is of the same
form as the cyclicity property of the bilinear form with respect to
the homotopies $b_n$  
\be
\label{cyclmn1}
\langle b_{n-1} (A_1, \ldots, A_{n-1}), A_n \rangle 
+ (-1)^{A_1 + \cdots + A_{n-1}} 
\langle A_1, b_{n-1} (A_2, \ldots, A_n) \rangle =0 \,,
\ee
where we have used (\ref{exchgpr}) to rewrite (\ref{cyclmn}).

If the homotopy associative algebra $\A$ (defined by the $b_n$)
satisfies the cyclicity condition (\ref{cyclmn1}), and the
homotopy map $f$ the condition (\ref{fcyclic}), it follows that the
homotopy associative algebra  $\widetilde \A$ (defined by the
$\tilde{b}_n$) satisfies the cyclicity condition (\ref{cyclmn1}). 
To see this, it is sufficient to show that for all $N$
\begin{eqnarray}
\label{checkzero}
& & \sum_{n=1}^{N-1} \sum_{l=1}^{n} 
\sum_{i_1+ \cdots + i_l=n}  \left[
\langle \tilde b_l \circ ( f_{i_1}\o \cdots \o f_{i_l} ) 
(A_1, \ldots , A_n), f_{N-n}(A_{n+1}, \ldots, A_{N}) \rangle \right. 
\\
& & \left. 
+ (-1)^{A_1 + \cdots A_{i_1}}
\langle f_{i_1}(A_1, \ldots, A_{i_1}), \tilde{b}_l\circ
(f_{i_2}\o\cdots\o f_{i_l}\o f_{N-n}) (A_{i_1+1}, \ldots, A_N) 
\rangle \right] \nonumber
\end{eqnarray}
vanishes. The desired identities are then recovered by a simple
induction argument. Using (\ref{gensete}), the first line can be
rewritten (after some relabeling) as
\be
\label{firstline}
\sum_{p=1}^{N-1} \sum_{l=1}^{N-p}
\langle f_l \circ b_p (A_1, \ldots, A_{l+p-1}),
f_{N-p+1-l} (A_{l+p}, \ldots, A_N) \rangle \,.
\ee
In the second line of (\ref{checkzero}), we relabel $m=i_1$, $r=N-n$,
and reassemble the sum over $l$ and $i_2, \ldots, i_l, r$ into an
expression of the form of the right hand side of (\ref{gensete}). Then
we can use (\ref{gensete}), and obtain (after some further
relabeling) 
\be
\label{secondline}
\sum_{p=1}^{N-1} \sum_{l=1}^{N-p}
(-1)^{A_1 + \cdots + A_{N-p+1-l}}
\langle f_{N-p+1-l} (A_1, \ldots, A_{N-p-l+1}), 
f_l \circ b_p (A_{N-p-l+2}, \ldots, A_N) \rangle \,.
\ee
The sum of (\ref{firstline}) and (\ref{secondline}) then vanishes
using (\ref{fcyclic}), and (for the case where $N-p-l+1=l=1$)
(\ref{invariant}) and (\ref{cyclmn1}). 

We can thus conclude that two open string field theories are
equivalent if the corresponding homotopy associative
algebras are homotopy equivalent, where the homotopy equivalence
induces an isomorphism of the underlying vector spaces which
preserves the bilinear form (\ref{invariant}), and the maps $f_n$
satisfy the cyclicity condition (\ref{fcyclic}).

\section{Tensor construction}
\setcounter{equation}{0}

Suppose that we are given a string field theory that is described in
terms of the basic structure, {\it i.e.} in terms of a homotopy
associative $*$-algebra $\A_\Cop$ with a suitable bilinear form
$\langle . , . \rangle$.  We want to analyze the conditions a set $\a$
has to satisfy so that $\A\o\a$ again possesses this basic structure.
We shall attempt to define one by one the various structures on
$\A\o\a$, and we shall, in each step, explain what conditions need to
be imposed on $\a$.

We shall also show that the tensor construction is {\it functorial}
in the sense that the two tensor theories $\A_1\o\a$ and $\A_2\o\a$ 
are strictly equivalent if the underlying open string field theories
based on $\A_1$ and $\A_2$ are strictly equivalent in the sense of
section~2.10.

\subsection{The vector space structure}

In this section we wish to explain that no generality can be gained
for the tensor product of the string field vector space $\A_\Cop$ with
the set $\a$ if we assume that $\a$ has less
structure than that of a vector space over $\Cop$.

Suppose first that $\a$ is a vector space over a field
$\F\subset\Cop$. We define the tensor product $\A_\Cop\o_\F\a$ to be
the vector space tensor product over $\F$, {\it i.e.} the quotient of the
direct product of sets $\A_\Cop \times \a$ by the identifications 
\begin{eqnarray}
(A+B) \o a & \sim & (A\o a) + (B\o a) \,,\nonumber \\
A \o (a+b) & \sim & (A\o a) + (A\o b) \,,\nonumber \\
(\lambda A) \o a & \sim &  A \o \lambda a \,,
\end{eqnarray}
where $A,B\in\A_\Cop, a,b\in\a$ and $\lambda\in\F$. This tensor product
is manifestly a vector space over $\F$. We can extend it to a vector
space over $\Cop$ using the property that $\A_\Cop$ is
a vector space over $\Cop$. 

We can also consider the complex vector space $\a_\Cop$
obtained from the vector space $\a$ by complexification, {\it i.e} by
choosing a basis and declaring the field to be $\Cop$. It is then easy
to see that $\A_\Cop \o_\F\a$ is isomorphic to the tensor product 
$\A_\Cop \o_\Cop\a_\Cop$. Thus, if $\a$ is a vector space we can assume
without loss of generality that the underlying field is $\Cop$.
\smallskip

If $\a$ is not a vector space, we define the ``tensor product''
$\A_\Cop\o\a$ to be the direct product of sets, and it is clear that
this set inherits the structure of a vector space over $\Cop$
from $\A_\Cop$. Alternatively, starting from $\a$,
we can define the complex vector space
$\a_\Cop$, whose basis is given by the elements of $\a$. It is
easy to see that the vector space tensor product  
$\A_\Cop \o_\Cop \a_\Cop$ is isomorphic to $\A_\Cop\o\a$. 
Thus, the set $\a$ can be replaced by the vector space $\a_\Cop$
without changing the tensor product. 

In the following we shall therefore assume, without loss of generality,
that $\a$ is a complex vector space, and we shall write
$\A\o\a$ for $\A_\Cop\o_\Cop\a$. This space is a vector space over
$\Cop$. It is clear that we can extend $\A\o\a$ to a module over the
appropriate Grassmann algebra $\G$, and that all the relevant properties
for this module will hold provided that they hold for $\A\o\a$. For
the time being, we shall not perform this extension explicitly, but we
shall be careful to include the signs which are to be relevant for the
extended module.
\medskip

We may assume that there exists a $\Zop$ grading on $\a$, {\it i.e.}
a map $\vare:\a\rightarrow\Zop$. (In particular, this grading may be
trivial.) We can then define a grading on $\A\o\a$, 
$\bvare: \A \o\a \rightarrow \Zop$, 
by 
\be
\label{dgrading}
\bvare(A\o a)=\vare(A) + \vare(a)\, . 
\ee
This gives $\A\o\a$ the structure of a $\Zop$-graded vector space
over $\Cop$. In the following it will be necessary to identify the
spaces  
\be
(\A\o\a)^{\o n} \simeq \A^{\o n} \o \a^{\o n} \,.
\ee
We shall use the conventions that 
\be
\label{identification}
\left( (A_1 \o a_1) \o \cdots \o (A_n \o a_n) \right) \simeq
(-1)^{s(a;A)} 
\left( A_1 \o \cdots \o A_n \right) 
\o \left( a_1 \o \cdots \o a_n \right)
\,,
\ee
where $s(a;A)$ is the sign which arises in the permutation of the
$a_i$ with the $A_j$, {\it i.e.}
\be
s(a;A) = \sum_{i=1}^{n-1} a_i \sum_{j=i+1}^{n} A_j \,.
\ee

\subsection{The BRST operator and the algebra structure}

In order to  define a BRST operator $\bQ$ on the tensor product
$\A\o\a$ we need to assume that there exists a linear map of degree
$-1$, $q:\a \rightarrow \a$ that squares to zero, {\it i.e.} $qqa=0$
for all $a\in \a$. For later convenience we shall write
$q(a)=m_1(a)$, and then the two properties are
\be
\label{feq}
m_1 \circ m_1 = 0 \,, \quad \vare(m_1(a))=\vare(a)-1\,.
\ee
We can then define $\bQ$ by  
\be
\label{ghty}
\bQ \equiv \bB_1 = b_1  \o \bbbone 
+  \bbbone \o m_1  \,.
\ee
This acts on $(A\o a)\in \A\o\a$ as
\be
\bB_1 (A\o a)  = (Q A \o a) 
+ (-1)^{A} (A \o q a) \,,
\ee
and one readily verifies that this is a differential, {\it i.e.} 
$\bB_1 \circ \bB_1=0$.
\medskip

To define a product on $\A\o\a$, we need to assume 
that there exists a product on $\a$, {\it i.e.} a bilinear
map $\a\o\a \rightarrow \a$ which we denote by  
$(a\o b) \mapsto m_2(a,b)$, and which is of degree zero
\be
\vare(m_2(a,b)) = \vare(a) + \vare(b)\,.
\ee
The product on $\A\o\a$ is then given as 
\be
\bB_2 = b_2 \,\o m_2 \,,
\ee
where we use (\ref{identification}) to identify $(\A\o\a)^{\o 2}$ with
$\A^{\o 2} \o \a^{\o 2}$. More explicitly, this product is
\be
\bB_2\left((A_1\o a_1),(A_2\o a_2)\right) = (-1)^{a_1 A_2}
\left(b_2(A_1,A_2)\o m_2(a_1 a_2)\right)\,.
\ee
The maps $\bB_1$ and $\bB_2$  will satisfy the second of the equations
(\ref{tryit}) provided that 
\be
\label{a1}
m_1 \circ m_2 - m_2 \circ m_1 = 0 \,,
\ee
where we use the convention that 
$m_2 \circ m_1 = m_2 \circ (m_1 \o \bbbone + \bbbone \o m_1)$. 
As explained in appendix~A, equations (\ref{feq}) and (\ref{a1}) are 
the first two of the family of identities which define a
homotopy associative algebra.

\subsection{Higher homotopies}

To define the higher homotopies $\bB_l$ for $l\geq 3$, it is
sufficient to assume that $\a$ has the structure of a
homotopy associative algebra. For our purposes it is most convenient
to assume that this structure is described in terms of the family of
maps $m_l:\a^{\o l}\rightarrow \a$, where $m_l$ is of degree $l-2$;
this formulation (which is the standard one for $A_\infty$ algebras)
is explained in detail in appendix~A. We shall not attempt to give a
full proof of the tensor construction here, but we want to give the
explicit expression for $\bB_3$ in the general case, and prove how the
construction works under the assumption that $\a$ is associative, 
{\it i.e.} $m_l=0$ for $l\geq 3$. 

In the general case where $\a$ is only homotopy associative, we define
\be
\label{bB3}
\begin{array}{lcl}
\bB_3 & = & {\displaystyle
\half b_3 \o \left(m_2 (m_2\o\bbbone + \bbbone\o m_2) \right) 
+ \half \left(b_2 (b_2\o\bbbone - \bbbone\o b_2)\right) \o m_3 \,,}
\end{array}
\ee
and it is not difficult to show that this expression (together with
the formulae for $\bB_1$ and $\bB_2$) satisfies (\ref{onvn}) for
$n=3$. We should mention that $\bB_3$ is not uniquely determined by
this condition. However, as shall be pointed out later, the above
expression is uniquely determined by the condition that the bilinear
form satisfies a certain cyclicity condition.

If the algebra $\a$ is strictly associative, we can give a closed
form expression for all the higher homotopies. We define for $l\geq 2$
\be
\label{bBl}
\bB_l = b_l \o \left(m_2 (m_2 \o \bbbone) \cdots 
(m_2 \o \bbbone^{\o (l-2)}) \right) \,.
\ee
This acts naturally on $\A^{\o l} \o \a^{\o l}$, and it defines a map
$(\A\o\a)^{\o l} \rightarrow (\A\o\a)$ using (\ref{identification}).

To show that this definition satisfies the conditions of a
homotopy associative algebra (\ref{onvn}),
we calculate for $l, n-l \geq 2$ 
\begin{eqnarray}
\bB_l \circ \bB_{n-l} & = &
\left(b_l \o \left(m_2 \cdots (m_2\o\bbbone^{\o(l-2)})\right) \right) 
\cdot \sum_{s=0}^{l-1} 
\left(\bbbone^{\o s} \o b_{n-l} \o \bbbone^{\o (l-1-s)}\right) 
\hspace*{2cm} \nonumber \\
& & \hspace*{1cm}
\o \left(\bbbone^{\o s} \o 
\left(m_2 \cdots (m_2\o\bbbone^{\o (n-l-2)})\right) \o
\bbbone^{\o (l-1-s)} \right)  \nonumber \\
& = & b_l \circ b_{n-l} \o 
\left( m_2 \cdots (m_2 \o \bbbone^{\o (n-3)}) \right)\,,
\end{eqnarray}
where we have used that $m_2$ is an even, associative product.
It then follows that
\begin{eqnarray}
\sum_{l=1}^{n-1} \bB_l \circ \bB_{n-l} & = &
\bB_1 \circ \bB_{n-1} + \bB_{n-1} \circ \bB_1
+ \sum_{l=2}^{n-2} \bB_l \circ \bB_{n-l} \nonumber \\
& = &
\left[ \sum_{l=1}^{n-1}
B_l \circ B_{n-l} \right] \o 
\left(m_2 \cdots (m_2\o \bbbone^{\o (n-3)})\right) =0\,.
\end{eqnarray}
This shows that the $\bB_n$'s define a homotopy associative algebra.

\subsection{Bilinear form}  

We define a bilinear form for the tensor product by 
\be
\label{bilinten}
\llangle\, (A\o a), (B\o b)\, \rrangle \equiv
(-1)^{a B} \langle A, B \rangle \; \langle a,b\rangle\,,
\ee
where $\langle\,  \cdot ,\cdot \rangle: \a\o\a\to \Cop$ 
is a bilinear form on $\a$. In order for this bilinear form
to be invertible on $\A\o\a$, we have to assume that 
$\langle \cdot ,\cdot \rangle$ is invertible on $\a$. 

The resulting bilinear form $\llangle \cdot,\cdot \rrangle$ has to be
odd and of fixed degree on $\A_\Cop\o\a$. Since 
$\langle \cdot,\cdot  \rangle$ is of odd degree on $\A_\Cop$, this
implies that $\langle \cdot,\cdot\rangle$ must be of even degree on
$\a$, {\it i.e.} that $\langle a, b \rangle$ vanishes unless the
degrees of $a$ and $b$ add to an even number. This is a very important
fact, and it is independent of the definition of $\bvare$ on $\A\o\a$,
as explained before at the end of section~2.2. It means, in
particular, that there is no simple way to tensor an open string
theory to itself. We shall assume, without loss of generality, that
$\langle \cdot,\cdot\rangle$ is of degree zero, 
\be
\label{degreem}
\langle a, b \rangle \neq 0 \quad \Rightarrow \quad
\epsilon(a)+\epsilon(b)=0 \,,
\ee
so that $\llangle \cdot,\cdot \rrangle$ is of degree $-1$.
We demand furthermore that 
\begin{eqnarray}
\label{a3}
\langle a,b\rangle & = & (-1)^{a b} \langle b,a\rangle\,, \\
\label{a4}
\langle qa,b\rangle  & = & - (-1)^{a} \langle a,qb \rangle\,,
\end{eqnarray}
and this will guarantee that 
$\llangle \cdot,\cdot \rrangle$ satisfies the exchange property 
(\ref{exchgpr}) and the $\bQ$-invariance
(\ref{qinnp}). In order for the bilinear form 
$\llangle \cdot,\cdot\rrangle$ to satisfy the cyclicity condition
(\ref{cyclmn}), it is sufficient to assume that 
\be
\label{cyclic}
\langle a_1, m_{n-1}(a_2, \ldots , a_n) \rangle = 
(-1)^{(n-1)(a_1 + a_2 +1) + a_1 (a_2 + \cdots + a_n)}
\langle a_2, m_{n-1}(a_3, \ldots , a_n, a_1) \rangle \,,
\ee
where, as before, the relevant signs are uniquely determined by the
consistency with the identities (\ref{mrels}). It is easy to check
that the cyclicity condition (\ref{cyclmn}) holds for $\bB_3$ 
(\ref{bB3}), and in fact, this condition fixes the ambiguity
in the definition of $\bB_3$. If $\a$ is strictly associative, the
cyclicity (\ref{cyclmn}) holds for $\bB_l$ (\ref{bBl}) for general
$l$.

\subsection{Star structure and reality}   

We can introduce a $*$-structure on $\A\o\a$ provided that there
exists a $*$-structure on $\a$, {\it i.e.} an antilinear map 
$a\mapsto a^\ast$, where $(a^\ast)^\ast = a$ and
$\vare(a^\ast)=\vare(a)$. In this case  we can define
\be
\label{starprod}
(A\o a)^\ast \equiv (-1)^{a (A+1)} A^\ast \o a^\ast\,.
\ee
We demand that  $\langle \cdot,\cdot \rangle$ satisfies
\begin{eqnarray}
\label{a5}
\overline{\langle a,b\rangle} & =  & \langle b^\ast,a^\ast \rangle \,, \\
\label{a6} (q a)^\ast & = & (-1)^{a} q a^\ast \,,
\end{eqnarray}
where the left hand side of the first line is the complex conjugation
in $\Cop$. It then follows that the bilinear form 
$\llangle \cdot, \cdot\rrangle$ satisfies (\ref{hermsym})
\begin{eqnarray}
\overline{\llangle\, (A\o a), (B \o b) \,\rrangle} & = &
(-1)^{a B} 
\overline{\langle a,b \rangle}\,
\overline{\langle A, B \rangle} \nonumber \\
& = & (-1)^{a B + (a+b)(A+B+1)} 
\langle B^\ast, A^\ast \rangle \langle b^\ast,a^\ast\rangle \nonumber \\
& = &
\label{tencon}
\llangle\, (B \o b)^\ast, (A\o a)^\ast \, \rrangle \,,
\end{eqnarray}
where we have used that the two forms in $\A$ and $\a$ are odd and
even, respectively. Likewise, the analogue of (\ref{qstar}) holds
\begin{eqnarray}
(\bQ (A\o a) )^\ast & = & (QA \o a)^\ast + (-1)^{A}
(A \o q a)^\ast \nonumber \\
& = & (-1)^{A(a+1)+1} (Q A^\ast \o a^\ast)
+ (-1)^{Aa+1} (A^\ast \o q a^\ast) \nonumber \\
\label{tenqst}
& = & (-1)^{(A\o a) +1} \bQ (A\o a)^\ast \,.
\end{eqnarray}

Having defined a star operation for $\a$, we can again impose the
reality condition $\Phi^* = \Phi$ in the tensor theory. This reality
condition will guarantee the reality of the string action.
A general string field in the tensor product is of the form
\be
\label{Phibasis}
\Phi= \sum_{ia} E_i e_a \phi^{ia} \,, 
\ee 
where $E_i\in\A_\Cop$ and $e_a\in\a$ are basis vectors (of definite
grade) of $\A_\Cop$ and $\a$, respectively, and $\phi^{ia}\in\G$. By
considering suitable linear combinations, if necessary, we may assume
that the vectors $E_i$ and $e_a$ are real, {\it i.e.} that 
$E_i^\ast = E_i$ and $e_a^\ast = e_a$. Using (\ref {starprod}),
(\ref{conjgr}) and the property that the string field is even, the
reality condition then becomes 
$\overline{\phi^{ia}} =(-)^{e_a(E_i+1)}\phi^{ia}$. In particular, the
target space fields associated to even elements of $\a$ are real.

\subsection{Equivalent tensor theories} 

In this section we shall show that the tensor construction is 
{\it functorial} in the sense that the tensoring of a given algebra
$\a$ into two strictly equivalent string field theories $\A$ and 
$\A'$ gives rise to two strictly equivalent string field theories
$\A\o\a$  and $\A'\o\a$. 

\smallskip
Let us suppose that we have two tensor product algebras, one with
homotopies 
\begin{eqnarray}
\label{thehom}
&&\bB_1 = b_1  \o \bbbone 
+  \bbbone \o m_1  \,\nonumber \\
&&\bB_l = b_l \o \left(m_2 (m_2 \o \bbbone) \cdots 
(m_2 \o \bbbone^{\o (l-2)}) \right) \quad l\geq 2\,,
\end{eqnarray}
and a second one with homotopies
\begin{eqnarray}
\label{thehoms}
&&\widetilde\bB_1 = \widetilde b_1  \o \bbbone 
+  \bbbone \o m_1  \,\nonumber \\
&&\widetilde\bB_l = \widetilde b_l \o \left(m_2 (m_2 \o \bbbone)
\cdots  (m_2 \o \bbbone^{\o (l-2)}) \right) \quad l\geq 2 \,,
\end{eqnarray}
where we have assumed, for simplicity, that the algebra described by
the $m$ maps is associative. Suppose further that the 
$b_n$'s and $\widetilde b_n$'s are related by a homotopy map, 
{\it i.e.} that there exist maps $f_n$ satisfying 
(\ref{gensete}). It is then easy to check that the
homotopy associative algebras $\bB$ and $\widetilde\bB$ are related
by a homotopy map which is defined by
\begin{eqnarray}
\label{themams}
&&F_1 = f_1  \o \bbbone \,\nonumber \\   
&&F_l = f_l \o \left(m_2 (m_2 \o \bbbone) \cdots 
(m_2 \o \bbbone^{\o (l-2)}) \right) \,, \,\, l\geq 2\,,
\end{eqnarray}
{\it i.e.} that the maps $\bB_n$, $\widetilde\bB_n$ and $F_n$ satisfy
the analog of (\ref{gensete}). If $f_1$ is an isomorphism of vector
spaces, so is $F_1$, and if the $f_n$'s satisfy the cyclicity
condition (\ref{fcyclic}), so do the $F_n$'s. 
\smallskip

Let us now assume that the underlying string field theory is
described in terms of the $b$ algebra. Then the above result implies
that the theory which is obtained by attaching an 
algebra $\a$
(described by the $m_n$ maps) to one formulation of the theory (in terms
of the $b_n$ maps, say) is equivalent to the theory, where 
the same  algebra
is attached to an equivalent version of the
underlying string field theory (described by the $\widetilde b_n$
maps). The tensor construction therefore does not depend on the
particular representation of the underlying theory.
\smallskip

We can also consider the situation, in which the underlying open
string field theory  $\A$
is cubic, and can thus be described in terms of
the $m$ maps. The above argument then implies that two
homotopy equivalent algebras $\a$ and $\a'$
(where the homotopy equivalence
gives rise to an isomorphism of vector spaces and satisfies the
cyclicity condition) give rise to equivalent string field theories
$\A\o\a$ and $\A\o\a'$.
This result is relevant for the classification of the tensor theories.

\section{Analysis of the tensor construction}
\setcounter{equation}{0}

We have now completed the construction of the tensor theory, and
discussed the ingredients that are necessary to build it. Next we
shall turn to analyzing these conditions in some detail. 
First, we shall explain how the positivity of the string field action
can be related to the property of a certain inner product to be
positive definite. We shall then analyze this inner product for the
tensor theory, where we assume that the additional degrees of freedom
do not have any spacetime interpretation. We shall find that the
tensor theory can only be positive if the cohomology of $\a$ is
concentrated at zero degree. Furthermore, the bilinear form must
induce a positive definite inner product on the cohomology $H(\a)$,
and this implies that $H(\a)$ is a semisimple algebra. We shall then
use a version of Wedderburn's theorem to show that $H(\a)$ is the
direct sum of a nilpotent ideal and a collection of full matrix
algebras over the complex numbers. 
\smallskip

In a second step we shall show that the string field theory
corresponding to $\A\o\a$ is physically equivalent to the string field
theory, where we replace $\a$ by the algebra  $H(\a)$, eliminating the
differential. As $H(\a)$ is a semisimple algebra over $\Cop$, this
will demonstrate that all the tensor theories we have constructed are
physically equivalent to the familiar Chan-Paton constructions.

\subsection{Positivity condition in open string theory}

In a quantum theory the condition of positivity requires that 
the set of physical states carries a positive definite inner product.
This requires, in particular, that an inner product is defined, and
therefore requires more structure than we have assumed so far. Indeed,
while we have a BRST operator that can be used to define physical
states, we only have an odd bilinear symplectic form and this cannot
be used to define a (symmetric) inner product. In a sense this is as
it should be, as an explicit positivity analysis of a field theory 
requires the examination of the propagators, and propagators only
exist after gauge fixing.

While no complete axiomatics of gauge fixing is known, it
is fairly clear how to proceed in specific situations. In
this paper it will suffice to analyze positivity explicitly for 
open bosonic strings, and for the Neveu-Schwarz (NS) sector of open
superstrings;
these two cases can be treated in essentially the same way.
A positivity analysis of the Ramond sector of open superstrings
would be interesting, but will not be necessary here.
\medskip

We can implement the gauge fixing condition by 
imposing a subsidiary condition which defines a  
subspace $\A_\Cop^{rel}$ of
$\A_\Cop$. The gauge fixed string field is an element of this
subspace, called the relative subspace,
and the physical states are determined by the BRST
cohomology calculated on $\A_\Cop^{rel}$; this is usually called the
relative cohomology $H^{rel}(\A_\Cop)$. The physical states also have
to satisfy a reality condition, and they 
can thus be identified with the 
real subspace of $H^{rel}(\A_\Cop)$. 

In the situations we are considering here, all physical states are
spacetime bosons, and this implies that the relative cohomology has to
be concentrated on even elements of $\A_\Cop^{rel}$. In this case, as
we shall explain below, the kinetic term of the string field theory
induces a natural symmetric bilinear form on the even subspace of
$\A_\Cop^{rel}$. It is then possible to rephrase the condition of
positivity as the condition that this symmetric bilinear form is
positive definite on the space of physical states. 
\smallskip

To gauge fix the open bosonic string (or the open NS superstring) we
choose the Siegel gauge \cite{Siegel}, which corresponds to defining 
$\A_\Cop^{rel}$ to consist of $\ket{\Phi}\in\A_\Cop$ which satisfy
$b_0 \ket{\Phi}=0$, where $b_0$ is the zero mode of the antighost
field. For such states the kinetic term becomes
\be
\label{Qc0}
\langle \Phi , Q \Phi \rangle \Bigl|_{ {b_0 \Phi =0} } \, = \,
\langle \Phi, c_0 L_0 \Phi \rangle \, ,
\ee
where $\{c_0,b_0\}=1$, and $\{Q,b_0\}=L_0$. It is obvious that this
kinetic term vanishes on the space of physical states.
This vanishing has a natural interpretation as
$L_0=0$ implements the mass-shell condition $(p^2 + m^2)=0$.
The propagator will therefore be positive provided
that the bilinear form
$\langle  \cdot , c_0 \cdot \rangle$, which is
obtained from (\ref{Qc0}) upon deleting $L_0$, is
positive definite and invertible
on the real subspace of $H^{rel}(\A_\Cop)$.

Indeed  $\langle  \cdot , c_0 \cdot \rangle$
is well defined on $H^{rel}(\A_\Cop)$\footnote{This demands that 
$\langle A , c_0 B\rangle$ for $A, B\in H^{rel}(\A_\Cop)$ must be
unchanged when $B\to B + Q \chi $, with $b_0\chi =0$. To check this
first note that $\chi = b_0 c_0 \chi$, and therefore we must show that
$\langle A , c_0 Q b_0 c_0 \chi \rangle$ vanishes. This is seen by
moving the $b_0$ towards the left until it annihilates $A$ ($b_0$ is
self adjoint with respect to the symplectic form). In moving $b_0$ we
pick an $L_0$ term by anticommutation with $Q$, which kills $A$, and
moving $b_0$ across $c_0$ we let $Q$ act directly on $A$, giving a
vanishing result. }.  Furthermore,
$\langle\cdot , c_0 \cdot \rangle$ defines a symmetric form
on $H^{rel}(\A_\Cop)$, since all elements in this space are even.  
It is therefore consistent to demand that
\be 
\langle \Phi, c_0 \Phi \rangle > 0 \,, \quad \mbox{for real}
\,\, \Phi \in H^{rel}(\A_\Cop) \,.
\ee  
In this case, the inverse will also be positive on the real
subspace of  $H^{rel}(\A_\Cop)$, thus giving rise to a 
positive propagator.

The above condition can be phrased more compactly
using the following result. Let $V_\Cop$ denote a complex vector
space  with a star operation; a bilinear form  
$\langle \cdot\,,\cdot\rangle$ is positive on the real subspace  if
and only if the associated sesquilinear form  
$\langle \cdot^*\, ,\cdot \rangle$ is positive on  all of $V_\Cop$.
Thus our requirement can be stated as the demand that the sesquilinear
form $( \cdot , \cdot )$ defined on $H^{rel}(\A_\Cop)$ as  
\be 
(A, B) \equiv \langle A^*\,,  c_0 B\rangle 
\ee 
is invertible, and positive, {\it i.e.}  $(H, H)>0$ for all 
$H\ne 0$, $H\in H^{rel}(\A_\Cop)$.

\subsection{Positivity condition for the tensor theory}

To analyze the positivity of the tensor theory, we shall now use the
same strategy as in the previous section. We will examine the positivity
constraints that arise when the underlying open string theory
is an open bosonic string or the NS sector of an open superstring
theory. It will be clear from the resulting constraints on the
internal sector that no further constraints would arise from the
Ramond sector of the open superstring theory.

In our analysis we will regard the degrees of freedom described by 
the vectors in $\a$ as internal, {\it i.e.} having no 
spacetime interpretation. This is an important assumption
as we will treat $\A$ and $\a$ on a different footing.
In particular, the relevant relative cohomology of the tensor theory
$H^{rel}(\A\o\a)$ will be defined as the cohomology of $\bQ$
on the subspace of states satisfying 
$(b_0\o \bbbone) \Phi = 0$. On this subspace 
we consider the bilinear form 
\be
\llangle \Phi_1, (c_0\o\bbbone) \Phi_2 \rrangle\,,
\ee
which arises from the spacetime part $(Q\o \bbbone)$ of the
BRST operator $\bQ$ in the kinetic term of the tensor theory. 
In analogy to the case considered before,
positivity will require the bilinear form to be positive
definite on the physical states, {\it i.e.} on the
real subspace of the relative cohomology $H^{rel}(\A\o\a)$. 
\smallskip

By the K\"unneth theorem, the relative
cohomology on the tensor product is  given as 
\be
H^{rel}(\A\o\a) = H^{rel}(\A_\Cop) \o H(\a) \,,
\ee
where $H^{rel}(\A_\Cop)$ is the relative cohomology defined before, and
$H(\a)$ is the cohomology of the internal differential $q$ on $\a$.
As before, the underlying open string theory we are considering
contains only  physical states that are spacetime bosons, and
therefore  $H^{rel}(\A_\Cop)$ contains only even elements. The algebra
$\a$ has no spacetime interpretation, and the physical states of the
tensor theory are therefore spacetime bosons as well. (For
example, they have a kinetic term appropriate for bosons, and
they transform in an integer spin representation of the Lorentz group.) 
As the overall string field must be even, this implies that $H(\a)$
must only contain even elements, as odd elements of $H(\a)$ would lead
to physical spacetime fields which anticommute, and therefore describe
fermions. The property that $H(\a)$ contains only even elements is
already a rather strong condition. We shall now show that even more
has to be true in order to satisfy the positivity condition: $H(\a)$
must be concentrated at degree zero.

Let $\Phi$ be in the physical subspace of the tensor theory, {\it i.e.}
$\Phi$ is a real vector in $H^{rel}(\A\o\a)$. We expand $\Phi$ as
\be
\Phi = \sum_{i,a} E_i  e_a \, \phi^{ia}\,,
\ee
where $E_i\in H^{rel}(\A_\Cop)$, and $e_a\in H(\a)$ are real, and
$\phi^{ia}\in\Cop$. As all vectors are even, the reality of 
$\Phi$ then translates into $\phi^{ia}\in\Rop$.
Positivity now demands that  
\begin{eqnarray}
\llangle \Phi, \left(c_0\o\bbbone\right) \Phi \rrangle & = &
\sum_{i,a,j,b} \llangle E_i \phi^{ia} e_a, \left(c_0\o\bbbone\right)\, 
E_{j} \phi^{jb} e_b \rrangle \nonumber \\
& = &
\sum_{i,a,j,b}\,\, \phi^{ia} \,
\langle E_i, c_0 E_j \rangle \,
\langle e_a , e_b \rangle \, \phi^{jb} \nonumber \\
& \equiv &
\sum_{i,a,j,b}\,\, \phi^{ia} \,
M_{ia , jb} \, \phi^{jb} 
\label{kincal}
\end{eqnarray}
is strictly positive for $\phi^{ia}\neq 0$. This is 
equivalent to the condition that the matrix $M$, defined above, is
positive. 

Using the exchange property (\ref{a3}) and the $Q$ invariance of
the bilinear form on $\A_\Cop$ we see that\footnote{Because of 
(\ref{Qc0}) and the property of $L_0$ to be hermitian and real, the
exchange properties of the bilinear form on $\A$ with respect to
$Q$ equal those with respect to $c_0$.}
\begin{eqnarray}
\langle E_i, c_0 E_j \rangle & = &  \langle E_j, c_0 E_i \rangle\,,
\nonumber \\ 
\langle e_a, e_b \rangle & = &  \langle e_b, e_a \rangle \,.
\end{eqnarray}
This implies that the matrix $M$ is symmetric under the
exchange of  $(i,a) \leftrightarrow (j,b)$. 
Furthermore, because of
\begin{eqnarray}
\overline{\langle E_i, c_0 E_j \rangle} & = &  
\langle E_i, c_0 E_j \rangle \,, \nonumber \\
\overline{\langle e_a, e_b \rangle} & = &  \langle e_a, e_b \rangle \,,
\end{eqnarray}
it follows that $M$ is real. 
Being symmetric and real, $M$
can thus be diagonalized. 

Suppose now that $H(\a)$ contains vectors of degree $m$, $m\neq 0$. As
the bilinear form on $\a$ couples only vectors of degree $m$ to
vectors of degree $-m$, we can decompose $H(\a)$ as
\be
H(\a) = H(\a)_0 \oplus H(\a)_{\pm} \,,
\ee
where $H(\a)_0$ contains the states at degree zero, and the
decomposition is into orthogonal subspaces. Likewise, we can decompose
$H^{rel}(\A\o\a)$ into orthogonal subspaces as
\be
H^{rel}(\A\o\a) = \left[H^{rel}(\A) \o H(\a)_0 \right] 
\bigoplus \left[ H^{rel}(\A) \o H(\a)_{\pm}\right] \,.
\ee
The condition that $M$ is positive on $H^{rel}(\A\o\a)$ is then
equivalent to the condition that $M$ is positive on each of the two
subspaces separately.  

If we choose a basis of vectors of definite degree for $H(\a)_{\pm}$,
the property that the bilinear form on $\a$ couples only vectors of
degree $m$ to vectors of degree $-m$ implies that $M$, restricted to 
$H^{rel}(\A) \o H(\a)_{\pm}$, vanishes on the diagonal. $M$ is
therefore traceless, and as $M$ can be diagonalized, it follows that
$M$ cannot be positive definite. We can therefore conclude that
$H(\a)_{\pm}$ must be empty, {\it i.e.} that $H(\a)$ only contains
states of degree zero. 

If this is the case, we can examine the positivity constraint
further. 
By virtue of the positivity of the original string field
theory, we can find a real basis $W_i$ for $H^{rel}(\A_\Cop)$ for
which 
\be
\label{ortho}
\langle W_i, c_0 W_j \rangle = \delta_{ij} \,.
\ee
Then (\ref{kincal}) becomes
\begin{eqnarray}
\llangle \Phi, \left(c_0\o\bbbone\right) \Phi \rrangle & = &
\sum_{i,a,j,b} \langle W_i , c_0 W_j \rangle \, \langle e_a, e_b \rangle
\varphi^{ia} \varphi^{jb} \nonumber \\
& = & 
\sum_{i} \left\langle \sum_{a} \varphi^{ia} e_a, \sum_{b} \varphi^{ib} e_b 
\right\rangle \,.
\end{eqnarray}
It is then clear that positivity is satisfied if and only if
the bilinear form on $\a$ is positive definite on the real subspace of
$H(\a)$. If we introduce the sesquilinear form  
\be
(a,b) \equiv \langle a^\ast, b \rangle \,,
\ee
then the condition becomes that $(\cdot,\cdot)$ defines a positive
definite inner product on $H(\a)$ (as was explained before at the end
of the previous section). We should mention that because of (\ref{a4})
and (\ref{a6}), this sesquilinear form is well-defined on $H(\a)$. 
\medskip

\subsection{Semisimplicity of $H(\a)$ and Wedderburn's
Theorem} 

It is well known that the cohomology $H(\a)$ of an $A_\infty$ algebra is
an associative algebra \cite{GJ}. Indeed, the product $m_2$ of $\a$
induces a well defined product on $H(\a)$ because 
$m_2 \circ  m_1 = m_1 \circ m_2$ guarantees  that the product of a closed
and a trivial element is trivial and thus zero in $H(\a)$. Moreover, 
the identity $m_1\circ m_3 + m_2 \circ m_2 + m_3 \circ m_1 =0$ implies
that the product on $H(\a)$ satisfies $m_2 \circ m_2 =0$, and 
thus that it is associative. 
The positivity condition  also implies that all elements of $H(\a)$
are at degree zero, and since the product $m_2$ is
of degree zero, $H(\a)$ is an ordinary associative algebra without
grading.  
Moreover, the induced bilinear form on $H(\a)$ is 
cyclic and the corresponding sesquilinear form positive definite.
We will now show that this implies that  $H(\a)$ is a semisimple
algebra.\footnote{Semisimplicity does not yet follow from the property
of the bilinear form to be invertible and cyclic. A finite-dimensional 
example is given by the $2$-dimensional associative (commutative)
algebra generated by $u$ and $w$ with relations $u ^2=w$, $uw=ww=0$
with the invertible symmetric (and cyclic) bilinear form given as
$\langle u, u \rangle =1$, $\langle u, w \rangle=1$, 
$\langle w,w\rangle=0$. This algebra is not semisimple, as the
subspace generated by $w$ is a subrepresentation, but there exists no
complementary subrepresentation.} 

Let us recall that we call an algebra $\B$ semisimple if the 
adjoint representation, {\it i.e.} the representation of $\B$, where
$a\in\B$ acts on $b\in\B$ by $a\cdot b = (ab)$, is completely reducible
into irreducible representations. (For a brief introduction into the
relevant concepts see for example \cite{FB}.) 

Suppose then that $\B$ contains a subrepresentation, {\it i.e.} a
subspace $\U\subset \B$, such that for every element $u\in \U$ and
$b\in\B$, $(bu)\in \U$. Because of the positive definite inner
product on $\B$ we can write $\B$ as a direct sum
\be
\B = \U \oplus \U^{\perp} \,,
\ee
where 
\be
\U^\perp \equiv \{ b\in \B : (u,b)=0\quad \mbox{for all $u\in \U$} \} \,.
\ee
We want to show that if $\U$ is a subrepresentation, then so is 
$\U^\perp$; by induction, we can then decompose the adjoint
representation completely, thereby showing that $\B$ is semisimple. 

To see that $\U^\perp$ is a subrepresentation, it is sufficient to show
that $(u,bv)=0$ for all $v\in \U^\perp$, $b\in\B$ and $u\in \U$. Using
the cyclicity of the bilinear form on $\B$, we have 
\begin{eqnarray}
(u,bv) & = & \langle u^\ast, b v \rangle  
=  \langle v, u^\ast b \rangle 
=   \langle u^\ast b , v \rangle  \nonumber \\  
& = & \langle (b^\ast u)^\ast, v \rangle  =  (b^\ast u, v)\,.
\end{eqnarray}
As $\U$ is a subrepresentation, $b^\ast u \in \U$, and thus the inner
product vanishes. This completes the proof.

We have thus shown that  $H(\a)$ is a semisimple associative algebra.
If $H(\a) $ is a finite-dimensional algebra, it follows from the analysis
of appendix~D that $H(\a)$ is a direct sum of a nilpotent ideal and a 
semisimple algebra with a unit element. By Wedderburn's theorem,  
also reviewed in appendix~D, the semisimple algebra with a
unit is a direct sum of finite dimensional matrix algebras over $\Cop$. 
We have therefore shown that 
\be
\label{mainreqxx}
H(\a) = I \bigoplus_{i} \M_{n_i}(\Cop) \, , \quad \epsilon (h) =0,
\,\, \hbox{for}\, h\in H(\a)\,, 
\ee 
where $I$ denotes the nilpotent ideal, and $\M_{n}(\Cop)$ denotes the
full matrix algebra of $n\times n$ matrices with complex entries.
It is easy to see that the nilpotent elements correspond to additional
copies of the open string theory, where all interactions have been
removed. 
In the following, we shall therefore mainly consider the case, where
$H(\a)$ is a direct sum of matrix algebras over $\Cop$. In this case
the bilinear form is the trace on each matrix algebra separately, and
the star operation corresponds to hermitian conjugation 
$A\mapsto A^\dagger$. (This is explained in appendix~E.) 
In order to distinguish this star operation form complex conjugation,
we shall denote the latter by an overline so that 
\be
\label{ccmatrix}
A^* \equiv  A^\dagger = \overline{A^t}\,,
\ee
where $A\in {\cal M}_n(\Cop)$. The ``real'' elements of the
complex matrix algebra, {\it i.e.} the elements satisfying $A^*=A$
are the hermitian matrices, and the ``imaginary" elements are the
antihermitian matrices.

\subsection{Physical equivalence of $\A\o \a$ and $\A\o H(\a)$ 
string theories}

In this subsection we want to show that the string theory defined on
$\A\o\a$, where $\a$ is an $A_\infty$ algebra satisfying the
conditions derived earlier, is {\it physically} equivalent to the
string theory defined on $\A\o H(\a)$, where $H(\a)$ is the
cohomology of $q$ on $\a$.
We say that two string field theories are physically
equivalent (at tree level) if there is a one-to-one identification of
physical states such that the tree level S-matrices coincide.  Two
physically equivalent string field theories may have completely
different sets of unphysical and gauge degrees of freedom.  

Physical equivalence is weaker than the strict off-shell equivalence
that holds when actions and symplectic forms are related by a
differentiable transformation. In this case the underlying vector
spaces are isomorphic. For physical equivalence, however, this is not
the case since in general $H(\a)$ is only isomorphic to a proper
subspace of $\a$.\footnote{We will not try to discuss here whether the
$A_\infty$ algebra $\a$ satisfying the conditions derived before, is
homotopy equivalent in our strong sense to a strictly associative
algebra. If so, any string field theory based on  an admissible
$A_\infty$ algebra would be {\it off-shell equivalent} to a string
theory based on an associative differential graded algebra. This
seems unlikely to be the case \cite{PSch}, 
but is an interesting question that could be answered 
using the theory of cyclic cohomology.}

In order to discuss the claimed equivalence we use a particular
decomposition of the vector space $\a$ into a direct sum of subspaces 
\be 
\label{decom}
\a = \a^p \oplus \a^e \oplus \a^u   \,,
\ee
where $\a^p$ and $\a^u$ can be identified (non-canonically) with the
cohomology classes of $q$ (the physical vectors), and
the unphysical vectors, respectively, and $\a^e$ is the subspace of
$q$-exact vectors in $\a$. (The unphysical vectors are those that,
except for the zero vector, are not annihilated by $q$.)
As discussed in appendix~C, we can choose $\a^p$ and
$\a^u$ in such a way that the bilinear form 
$\langle \cdot , \cdot\rangle$ on $\a$ only couples $\a^p$ to $\a^p$,
and $\a^e$ to $\a^u$. 
\smallskip

We consider the string field theory on  $\A\o\a$, and decompose
\be
\label{decalg}
\A\o\a = (\A\o\a^p) \oplus (\A\o\a^e )\oplus (\A\o\a^u)\,.
\ee
Correspondingly, we decompose the string field into
\be
\Phi = \Phi^p + \Phi^e+ \Phi^u\,,
\ee 
where $\Phi^p\in(\A\o\a^p)$, $\Phi^e\in(\A\o\a^e)$, and
$\Phi^u\in(\A\o\a^u)$. It should be stressed that we only decompose
with respect to the internal algebra; while $\Phi^p$ certainly
contains all physical states, as defined by the BRST operator $\bQ$ of
the tensor theory, it also contains vectors whose spacetime part are
not in the spacetime BRST cohomology. In fact, the string field
$\Phi^p$ is the  appropriate  string field for the string field theory
based on $\A\o\a^p$ or equivalently, based on  $\A\o H(\a)$.  
With some abuse of language, we will call $\Phi^p$, $\Phi^u$ and $\Phi^e$,
physical, unphysical, and exact, respectively. 
\smallskip

We will work with the cubic version of the underlying open string 
field theory,
but we shall not assume that the higher homotopies vanish on $\a$; as
a consequence, the string theory corresponding to $\A\o\a$ has
interactions at all orders.
Our objective will be to prove that
$\Phi^u$ and $\Phi^e$ lines never appear in tree-diagrams of
external $\Phi^p$ lines, and that the higher interactions are
irrelevant for such diagrams.
\medskip

We first examine the kinetic term of the string field theory on
$\A\o\a$. Using the above comments and recalling that 
${\bf Q} = Q\o \bbbone + \bbbone \o q$, we write 
\be
\label{kinten}
\half \langle \Phi , {\bf Q} \Phi \rangle = \half \langle \Phi^p
,Q\Phi ^p\rangle  +  \langle \Phi ^e, Q\Phi^u \rangle  + \half
\langle \Phi^u ,q\Phi ^u\rangle  \,.
\ee
The propagator is 
derived from the first two terms on the right hand side of
(\ref{kinten}) since those terms involve the spacetime  BRST
operator. We therefore have propagators coupling $\Phi^p$ to itself
and $\Phi^u$ to $\Phi^e$. The last term in the right hand side should
be thought of as a quadratic interaction coupling  $\Phi^u$ to itself.   
\medskip

Consider now a tree diagram for which all external lines are $\Phi^p$
lines, 
and denote by $N_{max}$ the 
maximal order of any vertex  in the
diagram. First we want to show that if such a diagram exists, then
there also exists a diagram with the same $N_{max}$, for which all
internal lines are either unphysical or exact. Indeed, 
if we are given such a diagram with internal
physical lines, we can cut the diagram along all physical lines to
obtain a number of tree diagrams whose external lines all belong to
$\Phi^p$, and which do not have any internal lines corresponding to
$\Phi^p$.  By assumption, however, at least one of these diagrams must
have a string vertex of order $N_{max}$. We shall now show that no such
diagram can be constructed.

Let $V^N (p_N, u_N, e_N)$ denote the number of vertices 
which have $N$ legs, $p_N$ of which are physical, $u_N$ unphysical and
$e_N$ exact. By construction we have
$p_N + u_N+ e_N = N$. For any tree diagram 
\be 
\label{newex}
E+ 2I = \sum_{N\geq 2} \sum_{p,u,e}  N V^N \, , 
\ee
where $E$ denotes the number of external legs, and $I$ denotes the
number of  internal legs. Since we are considering diagrams with no
loops, $I= V-1$, where $V$ denotes the total number of vertices. It
then follows that  
\be 
\label{newey}
E= 2+ \sum_{N\geq 2} \sum_{p,u,e}  (N-2)  V^N \,.  
\ee
Since the external legs are all physical, 
and all internal lines are either unphysical or exact, 
the number of external lines equals the number of physical
lines 
\be
E=\sum_{N\geq 3} \sum_{p,u,e} p_N\,  V^N \,, 
\ee
where we recall that $V^2$ does not appear here since it involves
two unphysical legs. Combining the last two equations we find that 
\be
2= \sum_{N\geq 3} \sum_{p,u,e}  \left[ p_N -N +2 \right] \, V^N \,. 
\ee
On the other hand, we can count the difference between the number
of unphysical and the number of exact legs in a diagram
\be
u-e = 2V^2 + \sum_{N\geq 3} \sum_{p,u,e}  \left[ u_N -e_N\right] \,
V^N \,. 
\ee
Finally, subtracting the last two equations we obtain
\be
\label{getrz}
u-e = 2 \Bigl[ 1 + V^2 + \sum_{N\geq 3} \sum_{p,u,e} (u_N-1)\, V^N \Bigr]\,.
\ee
As all internal legs are unphysical or exact, all propagators will
join an exact leg to a physical leg, and it follows that the number
$u-e$ must vanish.
\smallskip

Consider first the case, where all vertices have $N\leq 3$. In this
case (\ref{getrz}) becomes
\be
\label{getrzz}
u-e = 2 \Bigl[ 1 + V^2 + 2V_{uuu} + V_{uup} + V_{uue} \Bigr]\,, 
\ee
where 
$V_{uuu}$ denotes 
the number of three string vertices coupling three unphysical fields,
and similarly for the other vertices.  
The three string vertices with one unphysical leg do not contribute
because of the factor $u_3 - 1$. Furthermore, the vertices with no
unphysical legs are absent since $V_{ppe}, V_{pee}$, and $V_{eee}$
vanish by $q$-covariance, and $V_{ppp}$ is absent by the assumption
that there are no physical internal lines. As the right hand side of 
(\ref{getrzz}) is strictly positive, we cannot have $~u-e=0$, and we
have obtained the desired contradiction. 
We have thus shown that all internal lines are physical, and this
implies, in particular, that $V^2=0$.

Next, we consider the case when there are higher vertices in the
tree diagram. In this case, contributions such as $V_{eeee}$
need not vanish, and the above argument might fail. There is, however,
an additional constraint on a consistent diagram which is associated
with the internal degree. Indeed, the product $m_{N-1}$ has degree
$N-3$, and the interaction $V^N$ carries therefore a degree violation
of $(N-3)$. On the other hand, the interaction $V^2$ (which is of the
form $<\Phi^u , q \Phi^u>$) carries a degree violation of $-1$. As
the bilinear form and the propagator are of degree zero, the total  
degree violation of a diagram must be zero, and we have
\be
V^2 =  \sum_{N\geq 3} \sum_{p,u,e} (N-3)\, V^N \, . 
\ee
Inserting this equation into (\ref{getrz}) we obtain
\be
u-e = 2 \Bigl[ 1  + \sum_{N\geq 3} \sum_{p,u,e} (u_N + N-4)\, V^N \Bigr] \, .
\ee
The contributions for $N=3$ were shown  earlier to be all greater than
or equal to zero, and it is manifestly clear that the same holds for
$N\geq 4$. Consistent (degree preserving) diagrams therefore cannot
involve higher vertices. This is what we wanted to show.   
\medskip

In summary, we have shown  that the string field theory on $\A\o\a$
generates tree level diagrams that in the external $\Phi^p$ sector are
identical to those that would be generated by the action
\be
 \half \langle \Phi^p ,Q\Phi^p\rangle +  {1\over 3} 
\langle \Phi^p,   \Phi ^p \Phi^p \rangle \,.  
\ee
This is simply the string field theory action corresponding to 
$\A\o H(\a)$, and we have therefore established the claimed physical
equivalence.

\section{Twist symmetry, gauge groups and projections}
\setcounter{equation}{0}

So far we have analyzed the structure that needs to be imposed on the
algebra $\a$ in order for the tensor theory to have the basic
algebraic structure. We shall now consider the situation where the
original string field 
possesses a twist symmetry. We shall explain in detail how such a
symmetry operates on the various algebraic structures, and show that
it leaves the action invariant. We shall also show that the truncation
of the theory to twist eigenstates of eigenvalue one is consistent
with the master equation.

Using the twist symmetry of the theory we can identify the gauge group
of the tensor theory to be a direct product of $U(n)$ groups, where
each matrix algebra $\M_n(\Cop)$ gives rise to a factor of $U(n)$. We
can also extend the twist operator to a well-defined operator on the
tensor theory, and truncate to the eigenstates of eigenvalue plus
one. If $\a$ is a direct sum of matrix algebras, we classify all
inequivalent extensions of $\Omega$ to the tensor theory. We find that
the only allowed gauge groups which can be obtained from truncations
in this way are $SO(n)$, $USp(2n)$ and $U(n)$.

\subsection{Twist operator $\Omega$ }

Open string field theory may  possess a twist symmetry
related to the orientation reversal of the open string,
although, as far as we are aware, this additional symmetry is not
required by consistency. This symmetry is implemented by an even
linear operator $\Omega$ on $\A$ 
which satisfies
\be
\label{twbasic}
\epsilon (\Omega A) = \epsilon (A)\,, \quad
\Omega (A \lambda) = (\Omega A) \lambda\,.
\ee
In addition we require that it commutes with the BRST operator and
with the operation of star conjugation
\be
\label{qtwist}
\Omega \, Q = Q \Omega \,, \qquad 
\Omega \, *  = * \, \Omega \,.
\ee
This last property will ensure that we can impose simultaneously both
a reality condition and a twist condition on the string field. The
twist operator  must  also leave the symplectic form invariant 
\be
\label{twistsym}
\langle \Omega(A) , \Omega(B) \rangle = \langle A , B \rangle\, .
\ee
\smallskip

$\Omega$ acts on the higher homotopies by reversing the order of the
vectors and by acting on each vector separately. Furthermore, as
$\Omega^2$ corresponds geometrically to a trivial operation, it must
act as an automorphism $\tau=\Omega^2$ of the homotopy associative
algebra, {\it i.e.} 
\be
\tau b_n(A_1, \ldots, A_n) = b_n(\tau A_1, \ldots, \tau A_n) \,.
\ee
For bosonic theories this is automatically satisfied as
$\Omega^2=+1$, but $\tau$ may be non-trivial for fermionic theories.  
These requirements determine the action of $\Omega$ on the higher
homotopies as
\be
\label{signtw}
\Omega b_n ( A_1 , \ldots , A_n ) = (-)^{e_n(A)}  b_n ( \Omega A_n ,
\ldots , \Omega A_1 )\,. 
\ee
Here the sign factor $e_n(A_1, \ldots A_n)$ is the sign factor 
needed to permute (with signs) $A_1, A_2, \ldots, A_n$ into 
$A_n, A_{n-1},\ldots, A_1$, {\it i.e.} 
\be
\label{signtw1}
e_n(A_1, \ldots, A_n) = \sum_{p=1}^{n-1} \sum_{q=p+1}^{n} A_p A_q \,,
\ee
and the ambiguity is again a possible extra minus sign for even $n$.
The first two identities read
\begin{eqnarray}
\label{firsttwo}
\Omega b_2 (A_1, A_2) &=& (-)^{A_1A_2} 
b_2 (\Omega A_2 ,\Omega A_1) \nonumber\\ 
\Omega b_3 (A_1, A_2, A_3) &=& (-)^{(A_1+A_2) A_3 +A_1A_2 } b_3 
(\Omega A_3 ,\Omega A_2, \Omega A_1 ) \,.
\end{eqnarray}

{}From a more abstract point of view, as explained in appendix~B, 
the star operation 
(section~2.8)  and $\Omega$ are characterized by the
property that they act on products of vectors in the prescribed
way. In fact, up to some trivial redefinitions, there are only two
consistent sign conventions, and these correspond precisely to the
sign conventions of the star and the twist operation. 

The twist operator reverses the order of the vectors in a product, and
therefore a product of  twist eigenstates need not be a twist
eigenstate. However, we can introduce suitable linear combinations as 
\be
\label{commuta}
[ A , B]_\pm \equiv  AB \pm (-)^{AB} BA \,,
\ee
where $AB$ stands for $b_2(A,B)$.  It then follows from 
(\ref{firsttwo}) that
\be
\label{twcomm}
\Omega [A,B]_\pm = \pm [ \Omega A , \Omega B]\,.
\ee
If $A$ and $B$ are twist eigenstates of eigenvalues $\Omega_A$
and $\Omega_B$ respectively, then 
\be
\label{twistpr}
\Omega_{[A,B]_\pm} = \pm \Omega_A \,\Omega_B \, .
\ee
\smallskip

The twist symmetry of the open string theory implies that the
$SL(2, R)$ invariant vacuum $\ket{0}$ is odd under $\Omega$, 
{\it i.e.} $\Omega \ket{0} = -\ket{0}$. Indeed, we can use
$\Omega (A, B) = (-)^{AB} (\Omega B , \Omega A)$ with
$A=B=\ket{0}$ to obtain
\be
\Omega (\ket{0}, \ket{0}) = - (\Omega \ket{0} , \Omega \ket{0})\,.
\ee
Since the product of two vacuum states gives (in addition to other
states) the vacuum state itself, the above equation implies that
$\Omega \ket{0} = -\ket{0}$. The same applies obviously for the vacuum
state of a given momentum $p$, $\Omega \ket{p} = -\ket{p}$. 

In the bosonic string theory, $\Omega c_n \Omega= (-)^n c_n$, and 
$\Omega \alpha_n \Omega= (-)^n \alpha_n$. This implies that states of
the form $c_1\ket{p}$ associated to the tachyon field have 
$\Omega = +1$, while states of the form $c_1\alpha_{-1}\ket{p}$,
associated to the massless gauge fields have $\Omega = -1$. The same
conclusion also holds for the NS sector of the superstring theory.

As we shall explain in section~5.3, we can truncate the theory to the
states of $\Omega$-eigenvalue one. If we perform this truncation for
the bosonic string, we keep the tachyon and lose the gauge
field. While this does not seem to be physically interesting, it is
certainly consistent at tree level.

\subsection{Identification of the gauge groups}  

In order to be able to identify the gauge groups that arise in the
various constructions it is necessary to analyze the cubic term 
in the master action. This is the term of the form
\be
\label{three}
\llangle \Phi, \bB_2(\Phi, \Phi) \rrangle \,,
\ee
where $\Phi$ is the even string field satisfying the reality
condition $\Phi^\ast=\Phi$, and we assume that the internal algebra
is a matrix algebra over $\Cop$. If we expand $\Phi$ as before in 
(\ref{Phibasis}) we find after a simple manipulation
\be
\label{fghg}
\bB_2(\Phi,\Phi)  =  \sum_{i,a,j,b} 
\frac{1}{4} \Bigl\{ 
[E_i, E_j]_+\,
[e_a, e_b]_+  \,+\,  [E_i,E_j]_-\, 
[e_a, e_b]_- \Bigr\} \phi^{jb} \phi^{ia}\,,
\ee
where we have also used the commutator and anticommutator notation 
for the internal algebra, {\it i.e.} 
$[e_a,e_b]_\pm = e_a e_b \pm e_b e_a$, where
$e_a e_b = m_2(e_a , e_b)$. Then (\ref{three}) becomes
\be
\label{welldone}
\sum_{ia,jb,kc} (-1)^{E_k}
\frac{1}{4} \Biggl\{
\Bigl\langle E_k, 
[E_i,E_j]_+ \Bigr\rangle
\left\langle e_c, [e_a, e_b]_+ \right\rangle 
+ \Bigl\langle E_k,
[ E_i, E_j]_-\Bigr\rangle
 \left\langle e_c, [e_a,  e_b]_- \right\rangle
\Biggr\} \phi^{jb} \phi^{ia} \phi^{kc}\,.
\ee
The right hand side has two types of terms, one involving
the anticommutator and the other involving the commutator of the
internal algebra. We will now see that when the three spacetime
fields are massless vectors, only the commutator term survives.

It was shown in the previous subsection that a massless vector
corresponds to a state of $\Omega$-eigenvalue $-1$. The coupling of
three vectors is therefore determined by the cubic term, where
$E_i, E_j$ and $E_k$ are all eigenstates of $\Omega$ with  eigenvalue
$-1$. Using  the twist property (\ref{twistpr}) we find
\be
\Omega_{[E_i,E_j]_\pm} = \pm \Omega_{E_i} \Omega_{E_j} = \pm 1 \,.
\ee
Because of the invariance of the symplectic form on $\A$ under
$\Omega$, only the twist minus one state can couple to $E_k$. This is
the bottom state in the above equation, and corresponds to the second
term in equation (\ref{welldone}). The coupling of three massless
vectors is therefore proportional to  
\be
\label{gauge}
f_{abc} \equiv \left\langle e_c,
[e_a , e_b]_-  \right\rangle\,.
\ee
Letting $g_{ab} = \langle e_a , e_b\rangle$ denote the invertible
bilinear form with inverse $g^{ab}$, we see that the above equation
implies that  
\be
[ e_a , e_b]_{{}_-} =  f_{ab}^c\,  e_c \,, \qquad  \hbox{where}\qquad
   f_{ab}^c  = f_{abc} g^{cd}\,.
\ee
The numbers $f_{ab}^c$ define the structure constants of a Lie
algebra, the Lie algebra that is induced from the associative algebra
$\a$.  For the full matrix algebra $\M_n(\Cop)$ we obtain the complex
Lie algebra $gl(n,\Cop)$.  

As we must  restrict ourselves to real string fields, this is not the
resulting gauge algebra. If we write the string field as 
$\Phi = \sum_{i,a}  i E_i  (e_a \phi^{ia})$, 
where we have introduced the factor of $i$ for convenience, we
readily see that  reality of the string field requires that 
$(\sum_a e_a \phi^{ia})$ 
be an imaginary element of the matrix algebra (recall that the
$\phi^{ia}$ are complex numbers). If we  take the $\phi^{ia}$'s to be
real, then the $e_a$'s are purely imaginary, {\it i.e.} antihermitian
elements of $\M_n(\Cop)$. The antihermitian matrices in $\M_n(\Cop)$
generate the Lie-algebra $u(n)$, which is a real form of
$gl(n,\Cop)$.

\subsection{Twist invariance and truncations}  

In this section we use the properties of the twist operator $\Omega$,
postulated in  subsection~5.1, to show that the string field action
is twist invariant. It is then possible to truncate the classical
action consistently to the sector of the string field whose
$\Omega$-eigenvalue is $+1$.  

If the bilinear form is twist invariant and the products $b_n$
satisfy the twist property (\ref{signtw}), then
the master action will be twist invariant, {\it i.e.}
$S(\Phi) =S(\Omega\Phi)$. This is the case because each term in the
action is twist invariant
\be 
\langle \Omega \Phi \,, b_n (\Omega \Phi ,\ldots, \Omega \Phi ) \rangle
=\langle  \Omega \Phi \,, \Omega b_n (\Phi ,\ldots, \Phi ) \rangle
=\langle  \Phi \,, b_n ( \Phi ,\ldots ,  \Phi ) \rangle\,.
\ee
Here we have used the sign factor of (\ref{signtw1}) which
is $+1$ since the string field is even. Given our choice for the
action, the invariance of the action under $\Omega$ fixes thus
the sign ambiguity discussed in (\ref{signtw1}).

We can break the string field into eigenspaces of the twist
operator $\Omega$.  
Separating out the eigenspace of eigenvalue plus one, we write
\be
\label{twbreak}
\Phi = \Phi^+ +  \widetilde\Phi\,, \quad \hbox{where}\quad
\Omega\Phi^+ = + \Phi^+ \,, 
\ee
and $\widetilde\Phi$ represents the string fields whose twist 
eigenvalues are different from $(+1)$. 
Because of the invariance of the bilinear form under $\Omega$,
the symplectic form does not couple $\Phi^+$ to $\widetilde \Phi$.  
Furthermore, the action does not contain any terms which are linear in
$\widetilde \Phi$. The first property implies that the master
equation can be written as
\be
{\partial_r S \over \partial \phi^{i+}} \, \omega^{ij+}
{\partial_l S \over \partial \phi^{j+}} 
+{\partial_r S \over \partial \tilde\phi^\alpha } \, 
\omega^{\alpha\beta}
{\partial_l S \over \partial \tilde\phi^\beta}  = 0\, ,
\ee
where $\phi^{i+}$ are the target space fields associated to vectors
$E_i^+$  with twist eigenvalue plus one, and likewise for
$\tilde\phi^\alpha$. Since $S(\phi^{i+}, \tilde\phi^\alpha)$ is at
least quadratic in the tilde fields, the second term in the above
equation vanishes if we set $\tilde\phi^\alpha =0$. It thus follows
that $S( \phi^{i+}, \tilde\phi^\alpha =0)$ satisfies the equation 
\be
{\partial_r S(\phi^+ , 0) \over \partial \phi^{i+}} \, \omega^{ij+}
{\partial_l S(\phi^+, 0) \over \partial \phi^{j+}} =0 \,,
\ee 
and thus that we can consistently restrict the theory to the 
sector of twist-eigenstates with eigenvalue plus one. We should
mention that this reasoning only applies to the classical master
action, as the full quantum master equation contains second
derivatives of the action (with respect to the tilde fields) which
need not vanish when $\tilde\phi^\alpha =0$. The consistency of the
quantum theory is therefore not guaranteed after truncation, and must
be reexamined.  
\medskip

It will be of interest to consider the case when the open string
theory may have more than one kind of twist symmetry. For simplicity,
we focus  on the case where we have two commuting twist symmetries
$\Omega_1$ and $\Omega_2$,
\be
\label{twotwists}
\Omega_1 \Omega_2 = \Omega_2 \Omega_1 \,,
\ee
each of which squares to the identity. Let $P_{\Omega_i}$ denote the
projector to states with $\Omega_i = +1$. As explained before, the
invariance of the string action under $\Omega_i$ implies that the
truncated  action $S_i(\Phi)\equiv S(P_{\Omega_i}\Phi)$ satisfies the
master equation. It is then easy to see that $S_i$ is still
invariant under the second twist symmetry $\Omega_j$ (where $j\ne i$),
as 
\be
S_i (\Omega_j \Phi)= S (P_{\Omega_i}\Omega_j \Phi )
= S (\Omega_j P_{\Omega_i}\Phi ) = S (P_{\Omega_i}\Phi ) = S_i
(\Phi)\,. 
\ee
We can therefore truncate again to plus one eigenstates of $\Omega_j$,
and we conclude that $S(P_{\Omega_1} P_{\Omega_2}\Phi )$ defines
a consistent theory.

\subsection{Twist symmetry on $\A\o\a$}

Given a twist map $\Omega$ for $\A$, we can extend this to a twist map
for $\A\o\a$, provided there exists a linear twist map
$\omega:\a\rightarrow\a$ on $\a$. $\omega$ must commute
both with the internal differential $q$ and the star operation in $\a$
\begin{eqnarray}
\label{littleome}
\omega \, q &=& q\, \omega \,,  \nonumber \\
\, * \, \omega &=&  \omega \, * \, . 
\end{eqnarray}
Furthermore, $\omega$ must leave the bilinear form
$\langle \cdot,\cdot \rangle$ on $\a$ invariant
\be
\label{a8}
\langle \omega(a),\omega (b) \rangle = \langle a,b\rangle\,,
\ee
but we do not assume that $\omega^2$ is the identity. On the product,
$\omega$ must act as
\be
\label{a7}
\omega(ab) = (-1)^{a b} \omega(b) \omega(a) \,, 
\ee
in analogy with the first equation in (\ref{firsttwo}). 
\medskip

We can then define the twist operator $\bOmega$ on $\A\o\a$ as  
\be
\bOmega (A \o a) \equiv \Omega(A) \o \omega(a) \,.
\ee
It is straightforward to verify that $\bOmega$ commutes with $\bQ$ and
with the star operation of  $\A\o\a$ (the analogue of
(\ref{qtwist})). It also leaves the symplectic form
$\llangle \cdot , \cdot \rrangle$  invariant
(the analog of (\ref{twistsym})). Furthermore, in the case where $\a$
is associative, we also have (\ref{signtw})
$$
\bOmega \bB_n \left((A_1\o a_1), \ldots , (A_n \o a_n)\right)
\hspace*{8cm} 
\vspace*{-0.4cm}$$
\begin{eqnarray}
{}& = &
(-1)^{s(a;A)} 
\bOmega \left(b_n\left(A_1, \ldots, A_n\right) \o (a_1 \cdots a_n)
\right) \nonumber \\ 
& = & (-1)^{s(a;A) + s(A;A) + s(a;a)}
b_n\left(\Omega A_n, \ldots, \Omega A_1\right)
\o \left(\omega(a_n) \cdots \omega(a_1)\right) \nonumber \\
& = &
(-1)^{s(A\o a; A\o a)}
\bB_n\left(\Omega (A_n \o a_n), \ldots , \Omega (A_1 \o a_1)\right) \,,
\end{eqnarray}
where $s(a,A)$ is 
\be
s(a_1, \ldots, a_n; A_1, \ldots, A_n) = 
\sum_{l=1}^{n-1} a_l 
\left(A_{l+1} + \cdots + A_n\right)\,,
\ee
and likewise for $s(A;a)$, {\it etc}. 
Since the twist operator $\bOmega$ on $\A \o \a$ satisfies all the
required properties, the tensor open string theory is guaranteed to be 
invariant under twist.

\subsection {Twist operators for matrix algebras}

In this subsection we want to find all possible inequivalent ways in
which a linear twist map $\omega$ can be defined on $\a$. As explained
before, this will allow us to define a twist operator 
$\bOmega =\Omega \o \omega$ on $\A\o \a$, and will enable us to
truncate the resulting open string field theory consistently. We shall
take $\a$ to be the direct sum of full matrix algebras and a nilpotent
ideal. All elements of $\a$ are of degree zero, and the bilinear form
is defined by the trace.

In the present paper we shall restrict our considerations to the case 
where the automorphism $\Omega^2$ of the underlying string field theory 
is the identity. This is always the case for the bosonic
string theory, but there are interesting examples of fermionic
theories for which this does not hold. This will be discussed in some
detail elsewhere \cite{GZ2}.
\medskip

If $\Omega^2=+1$ on $\A$, the $\bOmega=+1$ eigenspace of the tensor
theory will only contain states in $\a$ for which $\omega^2=+1$. 
Indeed, the $\bOmega=+1$ subspace of the tensor product is of the form 
\be
\left(\A \o \a \right)_{\bOmega=+1} =
[ \A_{\Omega=+1} \o \a_{\omega=+1} \,] \, 
\oplus \, [ \A_{\Omega=-1} \o \a_{\omega=-1}]
\subset \A \o \a_0 \,,
\ee
where $\a_0$ is the subalgebra of $\a$ which is generated by all
elements of $\a$ which are invariant under $\omega^2$ 
\be
\a_0 = \{ a\in\a : \tau(a) = a \} \,.
\ee
As far as the resulting theory is concerned, it is therefore sufficient 
to tensor the subalgebra $\a_0$ of  $\a$.  
Because of the invariance properties of $\tau$, $\a_0$ is a
star algebra with a positive definite sesquilinear form, and thus
has the same structure that was assumed for $\a$.  We can thus assume,
without loss of  generality, that  on  the internal algebra $\a$, the
automorphism $\tau = \omega^2$ is the identity.   
\medskip

It is clear that on the nilpotent part of $\a$, $\omega$ can be any
(linear) invertible map $I\rightarrow I$. On a given matrix algebra
$\M_n (\Cop)$, (\ref{a7}) implies that $\omega(AB)=\omega(B)\omega(A)$
for $A,B\in \M_n(\Cop)$. It follows that the map  $j(A)=[\omega(A)]^t$
defines an automorphism of $\M_n$. Since all automorphisms of a full
matrix algebra are inner (see for example Ref.~\cite{Albert}, Theorem
5, p.~51), it follows that we can write 
\be 
\label{answ}
\omega_J(A) = J A^t J^{-1} \,, 
\ee 
where $A^t$ is the transpose of $A$ and $J$ is some invertible matrix
in $\M_n$. By construction, $\omega_J$ preserves the bilinear form
$\langle A, B\rangle = \hbox{tr}(AB)$ as demanded by (\ref{a8}). 

Not all different matrices $J$ generate inequivalent operators
$\omega_J$. If we consider a similarity transformation of the matrix
algebra generated by an invertible matrix $B$, the twist operator
$\omega_J$ becomes $\omega_{\tilde J}$ which is defined by
(\ref{answ}) with 
\be
\label{eqclass}
\widetilde J = B^t J B \,,  \quad  B \,\, \hbox{invertible} \,.
\ee
To parametrize the inequivalent twist operators we can therefore
consider the non-de\-ge\-ne\-rate bilinear form
$\langle \cdot , \cdot \rangle_J$ which is associated to $J$ by
\be
\langle v_1, v_2 \rangle_J \equiv v_1^t J v_2 \,.
\ee
Indeed, because of (\ref{eqclass}), the different twist operations
are in one-to-one correspondence with non-degenerate bilinear forms,
where we identify two forms if they are related by a transformation
of the underlying vector space.
\medskip

As has been discussed before, the automorphism $\tau=\omega^2$ is the
identity, and this implies that
\be 
\tau(A)  = \omega (\omega (A)) =   J (J^{-1})^t \, A \, J^t J^{-1} = A \, , 
\ee 
for all $A \in \M_n (\Cop)$.  It then follows from Schur's lemma that  
$J^t J^{-1}  = \lambda   \bbbone$, {\it i.e.} $J^t = \lambda J$. As the 
transpose operation squares to the identity, $\lambda=\pm 1$. We are
thus led to consider two cases, the case $J= J^t$, and the case 
$J= - J^t$. It is clear from (\ref{eqclass}) that these are
inequivalent cases, as the symmetry character of $J$ cannot be changed
by a similarity transformation. 
\smallskip

If $\lambda = + 1$, the bilinear form is symmetric and nondegenerate,
and we can successively find (complex) basis vectors so that $J$ is
the identity matrix in the corresponding basis\footnote{This is in
essence the Gram-Schmidt procedure: as the bilinear form is
non-degenerate, we can always find a vector whose inner product with
itself does not vanish. We can rescale the vector so that this inner
product is plus one. Then we consider the orthogonal subspace, and
continue.}. In this case, $\omega$ coincides with the usual
transpose of matrices, and a matrix $A$ satisfying $\omega(A)=-A$ is 
antisymmetric. 
\smallskip

If $\lambda=-1$, the bilinear form is antisymmetric and
nondegenerate. Since $\det(J^t)=\det(J)$, $J$ can only be invertible
if $J$ is a $2n\times 2n$ matrix, where $n\in\Nop$. In this case,
familiar in symplectic vector spaces, one can show (see, for example
\cite{Wood}) that there exists a basis in which $J$ is of the form 
\be
\label{J0}
J_0 = \left(
\begin{array}{cc}
0 & \bbbone \\
- \bbbone & 0 
\end{array}
\right) \,.
\ee
Then $J_0^{-1}=-J_0$, and a matrix $A$ which satisfies $\omega(A)=-A$,
{\it i.e.} $A=J_0 A^t J_0 $ is a {\it symplectic} matrix. 
A  $2n\times 2n$  (complex) symplectic matrix 
takes the form
\be
\label{s0}
A = \left(
\begin{array}{cc}
m  & n \\
p &  -m^t 
\end{array}
\right) \,, \quad n^t=n \,, \quad p^t = p\,,   
\ee
where  $m,n$ and $p$ are complex $n\times n$ matrices.  
\smallskip

This concludes the construction of the inequivalent twist operators for
a complex matrix algebra $\M_n$. We have found two inequivalent
operators for the case of even dimensional matrices, 
$\omega_I(A) = A^t$, and $\omega_{J_0} (A) = J_0 A^t J_0^{-1}$.   
For odd-dimensional matrices
only $\omega_I(A) = A^t$ is possible.

\subsection {$\bOmega$ projections}  

In this section we first consider the standard truncations
that give rise to orthogonal and symplectic gauge groups. We 
point out that the tensor theory corresponding to $\a=gl(2n,\Cop)$
has two commuting twist operators and this leads to the somewhat
surprising observation that an unoriented string theory may have a
nontrivial twist symmetry. We also consider the theory which is
obtained by imposing both twist projections, and we show that it is
equivalent to the $u(n)$ theory. 

We can restrict the open string theory to eigenstates of $\bOmega$
with eigenvalue $+1$. Then the states with $\Omega=+1$ have 
matrices with $\omega=+1$, and similarly, states with $\Omega=-1$ are
associated to matrices with $\omega=-1$. As expained in section~5.1,
the massless vectors have $\Omega=-1$, and correspond to the second
case.  

We may assume that all  matrices are  antihermitian (to guarantee
reality, as explained in section 5.2). Then
\begin{list}{(\roman{enumi})}{\usecounter{enumi}}
\item  For $\bOmega= \Omega \o \omega_I$, the  $\Omega= -1$
states are associated to matrices that are antisymmetric, in addition
to being antihermitian. Theses matrices are therefore real and
antisymmetric, and define the Lie algebra of $so(n)$. The states with
$\Omega=+1$ are associated to purely imaginary symmetric matrices. 
\item For $\bOmega= \Omega \o \omega_{J_0}$, the $\Omega =-1$ 
states are associated to matrices that  
are symplectic, in addition to being antihermitian, and thus define the
Lie algebra of $usp(2n)$. The states with $\Omega=+1$ are associated
to antihermitian matrices satisfying $A= -J_0A^tJ_0$.  
\end{list}

In the even dimensional case, there exist two inequivalent  twist
operators $\bOmega_I = \Omega\o \omega_I$ and  
$\bOmega_{J_0}= \Omega\o \omega_{J_0}$, which are easily seen to
commute. Following the discussion at the end of section 5.3,
we reach the somewhat surprising conclusion that  after orthogonal 
or symplectic projection, the resulting unoriented  
string theory still has a nontrivial twist symmetry.

We may truncate the theory to simultaneous eigenstates of $\bOmega_I$
and $\bOmega_{J_0}$. The resulting string theory then has two sectors
\be
\left[ \A_{\Omega=-1} \o  \a_{[ -1 ,  -1] } \right ]  \, \bigoplus \,\left[ 
\A_{\Omega=1} \o \a_{[ 1 , 1] } \right] \,,
\ee
where $\a_{[i,j]}$ denotes the subspace of $\a$ with $\omega_I = i$
and $\omega_{J_0} = j$. The gauge fields have $\Omega=-1$, and
therefore correspond to the $\a_{[-1,-1]}$ sector, which can be
identified as the space of $2n\times 2n$ matrices that are
antisymmetric and symplectic  
\be
\label{oneone}
M^-\in \a_{[-1,-1]}  \quad \leftrightarrow \quad  M^-  =  \left(
\begin{array}{cc}
a & s \\
-s & a 
\end{array}
\right) \,,
\ee
where $s$ and $a$ are  symmetric and antisymmetric,
respectively. The above matrices define a $n^2$ dimensional complex
vector space, and they generate precisely the Lie algebra of
$gl(n,\Cop)$. Indeed, the map that associates to the matrix in
(\ref{oneone}) the matrix $a+is$ is an isomorphism of Lie algebras. 
This suggests that the resulting theory is equivalent to the standard
$u(n)$ theory, that is obtained after the reality condition has been
imposed. In order to analyze this, we examine the 
$\Omega=+1$ sector, which is accompanied by matrices that are
symmetric and antisymplectic ({\it i.e.} have
$\omega_{J_0}=+1$)  
\be
\label{twotwo}
M^+\in \a_{[1,1]}  \quad \leftrightarrow \quad  M^+  =  \left(
\begin{array}{cc}
is & -ia \\
ia & is 
\end{array}
\right) \,,
\ee
where $s$ and $a$ are symmetric and antisymmetric,
respectively, and the factors of $i$ have been introduced for 
later convenience. These matrices define again an $n^2$-dimensional
complex vector space, and the two spaces $\a_{[1,1]}$ and
$\a_{[-1,-1]}$ are isomorphic. Indeed, if $M^+\in \a_{[1,1]}$ then 
$M^-=(-iJ_0 A)\in\a_{[-1,-1]}$, and vice versa as $J_0$ is invertible. 

It is not obvious, however, that this theory is really
equivalent to the tensor theory with $\a=gl(n,\Cop)$. In
particular, neither $\a_{[-1,-1]}$ nor $\a_{[1,1]}$ defines an
algebra, and the expected twist symmetry is hidden. Furthermore, the
vector space of the doubly projected theory does not even possess an 
algebra structure. However, as we shall now show, the string field
action of this theory is identical to the action of the tensor theory
with $\a=gl(n,\Cop)$, and therefore the two theories are equivalent. 
\medskip

In order to show that the two actions are the same, we choose a basis
of states for the $gl(n,\Cop)$ tensor theory of the form
$E_i^- \o g$ and $E_i^+ \o g$, where $E_i^\pm$ are basis states of
the underlying open string theory with twist eigenvalue $\pm 1$, and
$g\in \M_n(\Cop)$. We can always write $g\in \M_n(\Cop)$ as 
$g=a+is$, where $a$ is antisymmetric and $s$ is symmetric. 
We then define a map $f$, mapping states in the $gl(n,\Cop)$ tensor
theory to states in the double projected theory,
\begin{eqnarray}
\label{mapstheo}
f: \, E_i^- \o g  \mapsto E_i^- \o M^- (g) \,, &&\quad M^-(g) = 
\left(\begin{array}{cc} a & s \\ -s & a \end{array}\right)\nonumber \\
f:\, E_i^+ \o g  \mapsto E_i^+ \o M^+ (g) \,, &&\quad M^+(g) = 
\left(\begin{array}{cc} is & -ia \\ia & is \end{array}\right)\,.
\end{eqnarray}
This map is invertible, and acts trivially on the original open string
states. The kinetic terms of the two theories are identical, as the
identity 
\be
\label{trmap}
\hbox{tr} (g_1 g_2 ) = \half \,  \hbox{tr} \Bigl( M^\pm(g_1) M^\pm (g_2) 
\Bigr) \,, 
\ee
guarantees that the map is compatible with the symplectic structures
(the factor of one half is an irrelevant normalization factor).  
To analyze the cubic term 
$\llangle\Phi, {\bf B}_2 (\Phi,\Phi)\rrangle$ we determine the
commutator and the anticommutator of the $M^{\pm}(g)$ matrices, and
find  
\be
\label{Mg}
[M^{\eta_1}(g_1) , M^{\eta_2}(g_2) ]_\pm = M^{\pm\eta_1 \eta_2}
\Bigl([g_1,g_2]_\pm \Bigr)  \,.
\ee 
It then follows that  
\begin{eqnarray}
f(\bB_2(\Phi,\Phi)) & = &  {1\over 2}\sum_{ij} 
f\left([E^{\eta_i}_i g_i, E^{\eta_j}_j g_j]_+\right) \nonumber \\
& =  {1\over 2}& \sum_{ij}
\left( [E^{\eta_i}_i,E^{\eta_j}_j]_+ 
M^{\eta_i \eta_j} ([g_i,g_j]_+) 
+ [E^{\eta_i}_i,E^{\eta_j}_j]_-
M^{-\eta_i \eta_j} ([g_i,g_j]_-) \right) \nonumber \\
& = & [f(\Phi),f(\Phi)]_+ \,,
\end{eqnarray}
and this implies that the cubic term of the two theories is the same.
We may consider the case, where the original string field theory is
described in the associative Witten formulation, and we have therefore
shown that the two theories are equivalent. The twist symmetry of the
doubly projected theory corresponds to the transposition of the internal
matrix $g$, and therefore amounts to $a\to -a$.

The equivalence between the two descriptions of the $U(n)$ theory is
an example of an equivalence where the underlying algebraic structures
are considerably different. Indeed, the tensor theory with
$\a=gl(n,\Cop)$ possesses an $A_\infty$ algebra, whereas no such
structure seems to exist for the double projected theory. The example
therefore shows that the algebraic structure of an open string theory
cannot be recovered from the action. This is consistent with the idea
that open string theories may be dual to closed string theories, or
even membrane theories.

\medskip 

Finally, we consider the case, where 
$\a$ contains the direct sum of two or more isomorphic matrix
algebras $\M_n(\Cop)$. In this case, the algebra $\a$ has, in addition
to the inner automorphisms, outer automorphisms which permute the
isomorphic algebras $\M_n(\Cop)$. Since the automorphisms must square
to the identity, the permutation defining the outer automorphism must
act as a pairwise exchange of isomorphic algebras. It is therefore
sufficient to consider the case when we have only two isomorphic
algebras. If we write the matrices ${\bf A}$ of $\a$ in block-diagonal
form, then $\omega$ acts as 
\be
\omega({\bf A})=
\omega \left(
\begin{array}{cc}
A_1 & 0 \\
0 & A_2 
\end{array}
\right) = 
\left(
\begin{array}{cc}
\omega_0(A_2) & 0 \\
0 & \omega_0(A_1)
\end{array}
\right)\,,
\ee
where $\omega_0$ is either the ordinary or the symplectic
transposition on $\M_n(\Cop)$. For the $\Omega=-1$ sector relevant
to gauge fields, we impose the condition 
$\omega({\bf A})=-{\bf A}$ which restricts ${\bf A}$ to be 
of the form
\be
\left(
\begin{array}{cc}
A_1 & 0 \\
0 & -\omega_0(A_1)
\end{array}
\right)\,,
\ee
where $A_1$ is an arbitrary matrix in $\M_n(\Cop)$.  Reality of the
string field requires that $A_1$ is antihermitian, and the above
matrices generate thus the Lie algebra of $u(n)$. 
The $\Omega=+1$ sector is associated with matrices of the form
\be
\left(
\begin{array}{cc}
A_1 & 0 \\  
0 & \omega_0(A_1)
\end{array}
\right)\,,
\ee
which satisfy $\omega({\bf A})=+{\bf A}$. These matrices
are obviously in one-to-one correspondence to the ones satisfying 
$\omega({\bf A})=-{\bf A}$.
One can also verify that the resulting theory  has a twist symmetry  
$\bOmega = \Omega \o \widetilde\omega$, where $\widetilde\omega$  
acts on ${\bf A}$ as 
\be
\widetilde\omega ({\bf A}) =\widetilde\omega \left(
\begin{array}{cc}
A_1 & 0 \\
0 & A_2
\end{array}
\right) =
\left(
\begin{array}{cc}
\omega_0(A_1) & 0 \\
0 & \omega_0(A_2)
\end{array}
\right)\,.
\ee
This theory is also equivalent to the $u(n)$ theory; indeed, if
we define for $g\in gl(n,\Cop)$ 
\be
N^{-}(g) = 
\left(\begin{array}{cc}
g & 0 \\ 
0 & -\omega_0(g)
\end{array}
\right) 
\qquad
N^{+}(g) =
\left(\begin{array}{cc}
g & 0 \\ 
0 & \omega_0(g)
\end{array}
\right)\,,  
\ee
then the analogues of (\ref{trmap})  and (\ref{Mg}) hold.  
By the same arguments as before, this implies that the theory is
equivalent to the $u(n)$ theory.

\section{Summary and conclusions}

Let us briefly summarize our results. We have exhibited 
the algebraic structure of open string field theory as that
of an  $A_\infty$ algebra with an odd symplectic  
form, and we have analyzed the various conjugation operations in
detail. We have shown that this basic algebraic structure is
sufficient to guarantee that the action satisfies the classical master
equation of Batalin and Vilkovisky. We have also explained that two open
string field theories are equivalent if their homotopy associative
algebras are homotopy equivalent, where the homotopy equivalence
induces an isomorphism of the underlying vector spaces and satisfies a
cyclicity condition with respect to the bilinear form.

We have analyzed the structure of the set  $\a$ that can be  
consistently attached to an open string theory, and we have found that
\begin{list}{(\roman{enumi})}{\usecounter{enumi}}
\item in order to obtain a gauge invariant string field equation,
$\a$ has to be an $A_\infty$ algebra;   
\item in order for this string field equation to arise from an action
which satisfies the master equation, $\a$ has to have
an even invertible bilinear form with cyclicity properties;
\item in order for the action to be {\it real}, $\a$ has to have a
star structure with respect to which the bilinear form satisfies
natural properties;
\item when $\a$ has no spacetime interpretation, positivity requires  
that  the cohomology $H(\a)$ is concentrated at zero degree, and that
the sesquilinear form induced from the bilinear form  
and the star  involution is positive definite on $H(\a)$. This   
implies, in particular, that $H(\a)$ is a semisimple algebra.
\end{list}
In comparison to the analysis of Marcus and Sagnotti
\cite{MarSag} the present analysis differs mainly in that 
it is based on string field theory and the BV formalism, and that
much less structure about the internal vector space is assumed a
priori. For example we found that it is sufficient to demand that the
internal vector space has the structure of a homotopy associative
algebra $A_\infty$, rather than that of a strictly associative
algebra. The relevant tensor construction, associating to two
$A_\infty$ algebras a tensor $A_\infty$ algebra, does not
seem to have been considered before in the mathematical literature.
We have also given an explicit proof for the semisimplicity of the
algebra $H(\a)$ which arises as a consequence of the positivity of the
string field action.

We have shown that two tensor theories are equivalent if their
internal $A_\infty$ algebras are homotopy equivalent, where the
homotopy equivalence induces an isomorphism of the underlying vector
spaces, and satisfies a cyclicity property.  
It is known that there exist non-trivial homotopy associative algebras
which are not equivalent (in this sense) to associative algebras
\cite{PSch}, but we do not know whether there exist examples
which satisfy conditions (i)-(iv). It would be interesting to
analyze this question.

We have shown that every tensor theory for which $\a$ does not have
any spacetime interpretation is physically equivalent (in the
sense of having the same S-matrix elements) to the theory, where the
internal algebra $\a$ is replaced by its cohomology $H(\a)$. The
positivity of the string field action implies that this algebra is
semisimple, and we can use Wedderburn's theorem to conclude that 
$H(\a)$ is a direct sum of complete matrix algebras together with a
nilpotent ideal. This leads to a string theory whose gauge group is a
direct product of $U(n)$ groups.
\smallskip

More general tensor constructions are possible if the original open
string theory possesses additional structure. A typical example
(which we have discussed in this paper) is the case where the
original theory possesses a twist symmetry related to the orientation
reversal of the open strings. We can then consider truncations of the
theory, where we 
restrict the space of states to eigenstates of eigenvalue one of the
twist operator. In particular, we can extend the definition of the
twist operator to the tensor theory, and truncate the tensor theory in
this way.

For the case where the internal vector space is a direct sum of full
matrix algebras, we have classified the different definitions of a
twist operator for the tensor theory. Upon truncation, these
constructions lead to open string theories with gauge group $SO(n)$,
$USp(2n)$ and $U(n)$. We have noted the somewhat surprising fact that 
the unoriented theories with gauge group $USp(2n)$ and
$SO(2n)$  possess a non-trivial twist symmetry. These theories
can therefore be further truncated, but these truncations lead again to
gauge groups of the same type. 

In this analysis we have concentrated on the case where the twist
operator $\Omega$ of the original string field theory squares to the
identity operator. This is appropriate for bosonic string theories,
but does not seem to be necessarily the case for fermionic
theories \cite{GimPol}. In general, $\Omega^2$ only needs to be an
automorphism of the underlying $A_\infty$ algebra, and this allows for
more  general tensor constructions. Similarly, the theory may possess
additional symmetries, and further truncations are possible. Some of
these more exotic construction will be explained in detail in
\cite{GZ2}.
\medskip

The open string theories that are obtained by truncation from 
theories which are associated to $A_\infty$ algebras
do not typically possess this basic algebraic structure any
more. However, as we have seen in the case of the doubly projected
theory of section~5.6, the resulting theory may still be equivalent 
to a theory which can be described in terms of the basic
algebraic structure. This is an example of a more fundamental fact,
namely, that it is not possible to reconstruct the whole algebraic
structure from the action and the symplectic form. Indeed, some part
of the algebraic structure is ``redundant'', and does not play a role
in defining the action.

This is similar to the more familiar fact that it is impossible
to recover the action and the symplectic form from the S-matrix
elements of the physical states. An example for this fact that we have
discussed in this paper are the off-shell inequivalent theories
corresponding to $\A\o\a$ and $\A\o H(\a)$ that have the same physical
content at tree level. 

The situation is somewhat different in closed string theory, where
much, if not all of the algebraic structure ({\it i.e.} the 
homotopy Lie algebra) can be recovered from the action. This might be
related to the uniqueness of the closed string theory which is in
marked contrast to the richness of open string models (at the
classical level).
\medskip

One of the more fundamental lessons which can be drawn from our
analysis of the tensor construction is the fact that the internal
sector is necessarily of a structure that is
different from that of the
underlying (string) field theory. Indeed, as we have explained before,
the bilinear form of the internal vector space has to be even, whereas
the underlying field theory possesses naturally an odd bilinear
form. On the other hand, we have seen in the analysis of the
positivity of the action that an even bilinear form can be naturally
associated to a field theory as soon as we have fixed the gauge. This
seems to suggest that it might be possible to tensor a ``gauge-fixed''
field theory to a gauge invariant theory to obtain a gauge invariant
theory. The main complication with this idea is that we do not have
much a priori knowledge of the algebraic structures of gauge fixed
theories, and therefore, that it is not clear how to build such a
tensor theory in detail.

A possible application of this idea could be the construction of
a closed string field theory from two open string theories, where one
of the open string theories would be gauge invariant, and the other
one gauge fixed. In this construction the ``internal theory'' would
also have a spacetime interpretation, and the analysis of positivity
would have some new features. A proper understanding of
such tensor constructions should be important for our conceptual
understanding of field theory in general, and may lead to the
discovery of new nontrivial theories.

\section*{Acknowledgements}

\noindent  We thank  J.~Stasheff  for many helpful discussions, 
for explanations on homotopy equivalence, and for  
criticisms on a draft version of this paper. We are also
grateful to S.~Halperin, A.~Schwarz and C.~Vafa for 
illuminating conversations. Finally, we  thank M.~Markl,
P.~Goddard and T.~Lada for useful correspondences.

M.R.G. is supported by a  NATO-Fellowship and in part by NSF grant 
PHY-92-18167.
B. Z. is supported in part by funds
provided by  the U.S.~Department of Energy (D.O.E.) under contract
\#DE-FC02-94ER40818, and by a  fellowship of the
John Simon Guggenheim Memorial Foundation.

\appendix

\section{Homotopy associative algebras}
\setcounter{equation}{0}

It is useful to explain the relation between the standard presentation
of an $A_\infty$ algebra \cite{GJ}, and the one given in
section~2.7.  In essence, the two descriptions are related by a
{\it suspension} of the underlying vector space $V$. To explain this
in detail, let us denote by $V$ the vector space of 
section~2.7, and
let $W$ be the vector space, whose elements at degree $k-1$ are
precisely those of $V$ at degree $k$. The two vector spaces are
naturally isomorphic, and the isomorphism which takes the elements of
$W$ at degree $k-1$ to those of $V$ at degree $k$ is called the
suspension map $s: W \rightarrow V$.

The multilinear maps $b_n : V^{\o n} \to V$ induce then naturally
multilinear maps $m_n : W^{\o n} \to W$, where
$m_n = s^{-1} b_n s^{\o n}$. More explicitly, this means that
\be
\label{btom}
s^{-1} b_n  (sA_1, \cdots sA_n ) = 
(-)^{s (A)} m_n (A_1 , \cdots A_n ) \,, 
\ee
where $s(A) = (n-1) A_1 + (n-2) A_2 + \cdots + A_{n-1}$. As all
$b_n$'s have degree $-1$, it follows that the degree of $m_n$ 
is $n-2$. The first few cases of eqn.(\ref{btom}) give 
\begin{eqnarray}
\label{ffew}
s^{-1} b_1  (sA)  &=& m_1(A)\,, \nonumber   \\
s^{-1} b_2 (sA, sB) &=& (-)^A m_2 (A, B)\, \nonumber  \\
s^{-1} b_3 (sA, sB, sC) &=& (-)^B m_3 (A,B,C) \, . 
\end{eqnarray} 
Because of the nontrivial signs, the identities of the
homotopy-associative algebra involving the $b_n$ maps are somewhat
different from those involving the $m_n$ maps. For example, the second
equation in (\ref{tryy}) can be rewritten as
\begin{eqnarray} 
\label {transl}
0&=& b_1 \circ b_2 (sA, sB)  + b_2 ( b_1 (sA), sB)  + (-)^{A+1} b_2 (
sA, b_1 (sB)\, )  
\nonumber \\
&=& s\left((-)^A m_1 \circ m_2 (A, B)  + (-)^{A-1} m_2 ( m_1 (A), B)  
- m_2 ( A, m_1 (B)\, )  \right) 
\nonumber \\
&=& (-1)^A s \left(
m_1 \circ m_2 (A, B)  - m_2 ( m_1 (A), B)  - (-)^A m_2 ( A, m_1
(B)\, )\right)  \,.
\end{eqnarray}
Letting $m_1 (A) = QA$ and $m_2 (A, B) = (A,B)_m$, this last identity
reads 
\be
\label{readb}
Q (A,B)_m =   (QA, B)_m   + (-)^A  (A,QB)_m \,,  
\ee 
which is recognized as the statement that $Q$
is an odd derivation of the (even) product $(\,, \, )_m$, with
conventional sign factors. In more generality, the $m$-identities are
given as \cite{PSch,Lada}
\be
\label{mrels}
\sum_{l=1}^{n-1} (-1)^{n(l+1)} m_l \circ m_{n-l} = 0\,,
\ee
where we use the notation that
\be
m_p = \sum_{r=0}^{n-p} (-1)^{r (p+1)} \bbbone^{\o r} \o
m_p \o \bbbone^{\o (n-p-r)}
\ee
on $V^{\o n}$. The first few of these identities read
\be
\label{trym}
\begin{array}{lllcl}
\hbox{On} \, W & : \quad &   0 &=& m_1 \circ m_1  \,, \nonumber \\
\hbox{On}  \, W^{\otimes 2} & : & 0 &=& 
m_1 \circ m_2 - m_2 \circ (m_1 \otimes \bbbone 
+ \bbbone \otimes m_1 ) \,,  \nonumber \\
\hbox{On}  \, W^{\otimes 3} & : &  0 &=& m_1 \circ m_3 + m_2 \circ
(m_2 \otimes \bbbone - \bbbone \otimes m_2 ) \,,  \nonumber \\
& & & & \quad + m_3 \circ ( m_1 \otimes \bbbone\otimes \bbbone
+ \bbbone \otimes m_1 \otimes \bbbone +  \bbbone \otimes
\bbbone\otimes m_1) \,.
\end{array}
\ee
The last identity can be written as
\begin{eqnarray}
\label{convfr}
((A, B)_m , C)_m -  (A,(B,C)_m)_m  &=&  - Q(A,B,C)_m   -  (QA,B,C)_m
 \\ 
&{}&  - (-)^A (A,QB,C)_m  - (-)^{A+B} (A,B,QC)_m\,, \nonumber
\end{eqnarray}
which is recognized as the violation of associativity.

\section{Systematics of signs for star and twist}  
\setcounter{equation}{0}

In this appendix we want to explain how the signs for the operations
$*$ and $\Omega$ can be understood from a more fundamental
point of view. In particular, we want to show that there exist (in
essence) only two operations (namely $*$ and $\Omega$) which are 
well-defined on $T(V)$, and which satisfy natural relations with
respect to the comultiplication $\Delta:T(V)\rightarrow T(V)\o T(V)$
and the coderivation ${\bf b}:T(V)\rightarrow T(V)$. 
\smallskip

We want to analyze the class of maps on $T(V)$ which are of the form
\be
\bj(A_1 \o \cdots \o A_n) = (-1)^{\epsilon_{j,n}(A_1, \ldots, A_n)}
\left(j(A_n) \o \cdots \o j(A_1) \right) \,,
\ee
where $j$ is a linear map $j: V \to V$ of degree zero, and
$\epsilon_{j,n}(A_1, \ldots, A_n)$ are some numbers which we
want to determine. By definition, $\bj: T(V) \to T(V)$ is of degree
zero, and $\epsilon_{j,1} = 0$. 
If we demand that 
$\bj^2$  is an automorphism, {\it i.e.} that 
\be
\bj^2(A_1 \o \cdots \o A_n) = \left(j^2(A_1) \o \cdots \o 
j^2(A_n) \right) \,,
\ee
this implies that
\be
\label{ordertwo}
\epsilon_{j,n}(A_1, \ldots, A_n) + \epsilon_{j,n}(A_n, \ldots, A_1)
= 0 \qquad \mbox{mod}\, 2\,.
\ee
Next, we analyze the condition that 
$\bj$  be compatible with the comultiplication $\Delta$, {\it i.e.}
\be
\bj \circ \Delta = \Delta \circ \bj \,.
\ee
A simple calculation then gives that $\epsilon_{j,n}(A_1, \ldots, A_n)$
has to satisfy the following set of equations (mod $2$)  
\begin{eqnarray}
\epsilon_{j,n}(A_1, \ldots, A_n)  & = &
\epsilon_{j,r}(A_1, \ldots, A_r) 
+ \epsilon_{j,n-r}(A_{r+1}, \ldots, A_n) \nonumber \\
\label{recurs}
& & \qquad 
+ \epsilon_{j,2}(A_1+ \cdots + A_r, A_{r+1} + \cdots+ A_n)\,,
\end{eqnarray}
where $r=1, \ldots n-1$. By definition, $\epsilon_{j,1}(A)=0$, and the
first non-trivial condition arises for $n=3$, {\it i.e.}
\begin{eqnarray}
\label{recurs1}
\epsilon_{j,3}(A_1,A_2, A_3) & = & 
\epsilon_{j,2}(A_1,A_2) + \epsilon_{j,2}(A_1+A_2,A_2) \nonumber \\
& = & 
\epsilon_{j,2}(A_2,A_3) + \epsilon_{j,2}(A_1,A_2+A_3) \,.
\end{eqnarray}
If we make the ansatz that 
\be
\epsilon_{j,2}(A,B) = \alpha A + \beta B + \gamma A B + \delta \,,
\ee
we find that $\alpha=\beta$ (from (\ref{ordertwo})), and that 
$\alpha=0$ from (\ref{recurs1}). By redefining $\bj$ by $-\bj$, if
necessary, we may also assume that $\delta=0$. We are thus left with
two possible solutions
\begin{eqnarray}
\epsilon_{1,2}(A,B) & = & 0 \label{sol1} \\
\epsilon_{2,2}(A,B) & = & A B\,. \label{sol2}
\end{eqnarray}
It is also clear that $\epsilon_{j,n}$ for $n\geq 3$ is determined by 
$\epsilon_{j,n}$ for $n=1,2$, as follows directly from (\ref{recurs}).

Next we want to assume that $j$ satisfies a natural relation with
respect to the coderivation, {\it i.e.} that
\be
\bj \left({\bf b}({\bf A}) \right) = (-1)^{\delta_j({\bf A})}
{\bf b} \left(\bj ({\bf A}) \right) \,,
\ee
where ${\bf A}\in T(V)$, and $\delta$ can only depend on the grade of
${\bf A}=A_1\o \cdots \o A_n$ in $T(V)$, {\it i.e.} the sum of grades
of $A_i$. Let us first analyze the consistency condition which arises
from the evaluation on ${\bf A}=A_1\o A_2$. A short calculation gives
then the conditions (mod $2$) 
\begin{eqnarray}
\epsilon_{j,2}(b_1 A_1, A_2) + \delta_j(A_1) & = &
\delta_j(A_1 + A_2) + A_2 + \epsilon_{j,2}(A_1,A_2) \nonumber \\
A_1+ \epsilon_{j,2}(A_1, b_1 A_2) + \delta_j(A_2) & = &
\delta_j(A_1 + A_2) + \epsilon_{j,2}(A_1,A_2) \,. \nonumber 
\end{eqnarray}
For the first solution for $\epsilon_j$ (\ref{sol1}), we then find the
recursion relation $\delta_1(A) + B = \delta_1(A+B)$, and for the
second solution (\ref{sol2}), we find $\delta_2(A) =
\delta_2(A+B)$. This fixes $\delta$ up to an overall constant
\begin{eqnarray}
\delta_1(A) & = & A + \rho_1 \label{sol1d} \\
\delta_2(A) & = & \rho_2\,. \label{sol2d}
\end{eqnarray}
The first solution (\ref{sol1},\, \ref{sol1d}) corresponds to the star
operation (where $\rho_1=1$), and the second solution (\ref{sol2},\,
\ref{sol2d}) corresponds to the twist operation (where
$\rho_2=0$). The constants $\rho_j$ are not fixed by consistency; in
the first case, this choice corresponds to whether  the BRST
operator $Q$ is hermitian ($\rho_1=1$) or antihermitian
($\rho_1=0$), and in the second case whether the twist commutes
($\rho_2=0$) or anticommutes ($\rho_2=1$) with the BRST operator.

\section{Physical, trivial and unphysical states}
\setcounter{equation}{0}

Suppose we are given a vector space of states $\H$, together with an
invertible bilinear $\langle\cdot,\cdot\rangle$ form on $\H$ for
which 
\be
\langle \psi, \chi \rangle = (-1)^{\varepsilon}
\langle \chi, \psi \rangle \,,
\ee
where $\varepsilon=0,1$, depends on $\psi$ and $\chi$. Let us further
assume that $\H$ carries an action of a BRST operator $Q$ which
squares to zero and for which 
\be
\label{qcov}
\langle Q\psi,\chi\rangle = (-1)^{\varepsilon_1}
\langle \psi, Q \chi\rangle\,,
\ee
where $\varepsilon_1=0,1$ depends on $\psi$ and $\chi$. $\H$
contains the subspace of closed states $\C\subset\H$, where $\C$
contains all states $\psi\in\H$ for which $Q\psi=0$. Because of
$Q^2=0$, $\C$ contains the subspace $\E$ of exact states, {\it i.e.}
the states of the form $Q\chi$ for some $\chi\in\H$. The quotient
space $\C / \E = H(\H)$ is called the cohomology of $Q$. 
Because of (\ref{qcov}) the bilinear form  on $\H$ induces a bilinear
form on $H(\H)$. We assume that this induced form is invertible.  

A priori, the quotient space $H(\H)$ of $\C$ representing physical
states does not have a canonical identification as a subspace of $\C$,
and likewise, the quotient space $\H / \C$ representing unphysical
states does not have a canonical identification as a subspace of
$\H$. However, we can always    
find subspaces $\P\simeq H(\H)$ and $\U\simeq \H / \C$ of $\H$ so that
$\P\subset \C$ and 
\be
\H = \E \oplus \P \oplus \U = \C \oplus \U \,,
\ee
where the direct sum $\oplus$ means that the pairwise intersection of
any two of the spaces contains only the zero vector. We shall refer to
$\P$ and $\U$ as the {\it physical} and {\it unphysical} subspace,
respectively, and we shall call $\E$ the {\it trivial} subspace.  
Note that being a subspace, $\U$ contains the zero vector, but
any non-zero vector in $\U$ is not annihilated by $Q$ (if it were it
would have to be in $\C$).  

\medskip
We want to show in this appendix that $\P$ and $\U$ can be chosen in
such a way that the bilinear form defines a matrix 
\be
\left( \begin{array}{ccc}
0 & 0 & N \\
0 & M & 0 \\
\hat{N} & 0 & 0 
\end{array}
\right)\,,
\ee
where we have written the matrix in blocks corresponding to $\E$, $\P$
and $\U$. First of all, because of (\ref{qcov}), the matrix elements
of states in $\E$ with states in $\C$ vanish automatically. This
implies that the upper $2\times 2$-block matrix has the structure we
demand, irrespective of the choice of $\P$. 

Next we want to modify $\U$ (leaving $\P$ unchanged) so that the
matrix elements of $\P$ and $\U$ vanish. Let us choose a basis $u_i$ 
for $\U$. Each $u_i$ defines a linear functional on $\P$ by 
\be
\alpha_i(p) = \langle u_i, p \rangle \,,
\ee
where $p\in\P$. As the restriction of the bilinear form to $\P$ is
invertible, we can then find an element $p_i\in\C$ so that 
\be
\alpha_i(p) = \langle p_i, p \rangle \,, 
\ee
for all $p\in\P$. Then we define $\hat{u_i}=u_i-p_i$, and by
construction, $\hat{u_i}$ is orthogonal to $\P$. We denote by $\U$
the space which is spanned by the basis $\hat{u}_i$. 

Next we want to show that we can modify the basis $\hat{u}_i$ further
(by suitable addition of elements in $\E$) so that the matrix element
$\langle \hat{u}_i, \hat{u}_j \rangle$ vanishes for all $i$ and
$j$. As we are adding elements in $\E$, this does not modify the
property that the $\hat{u}_i$ are orthogonal to $\P$. 

Let $u$ be an arbitrary (non-trivial) element of $\U$. Then, as $u$ is
not contained in $\C$, $Qu\neq 0$, and since the bilinear form is
non-degenerate, there exists an element $Q\chi \in \E$ so that   
\be
\langle Q \chi, u \rangle = (-1)^{\varepsilon}
\langle \chi, Q u \rangle \neq 0\,,
\ee
where $\chi\in\U$. It is clear that $Q$ defines a bijection of $\U$
onto $\E$, {\it i.e.} that every element in $\E$ is the image of a
suitable element in $\U$ under $Q$, and that different elements in
$\U$ are mapped to different elements in $\E$. As the bilinear form is
invertible, it then follows that every linear functional on $\U$ can
be written in terms of a vector in $\E$.   

We can now successively redefine
the vectors $\hat{u}_i$ in order to make all their inner products
vanish. Assume we have obtained a series of vectors $\tilde{u}_k$,
with $k<n$ all of whose inner products vanish. To obtain 
$\tilde{u}_{n}$ we proceed  as follows. We subtract from $\hat{u}_n$
a suitable trivial vector so that  the resulting vector $\hat{u}'_n$
satisfies  $\langle \hat{u}'_n,  \hat{u}'_n \rangle = 0$.  Then we add
further trivial vectors so that the resulting vector $\tilde{u}_n$
satisfies $\langle \tilde{u}_n,  \tilde{u}_k\rangle =0$, 
for $k< n$, and $\langle  \tilde{u}_n,  \hat{u}'_n \rangle =0$.  
It then follows that $\tilde{u}_n$ has vanishing inner product
with itself, and therefore vanishing inner product with all
$\tilde{u}_j$ for $j\leq n$, as desired. By recursion we can 
thus find a basis $\tilde{u}_i$ so that
\be
\langle \tilde{u}_i, \tilde{u}_j \rangle = 0  \,,
\ee
for all $i,j$. This concludes our proof.

\section{Wedderburn theory}
\setcounter{equation}{0}

Let us briefly summarize the results which we need for our discussion.
The standard Wedderburn theorem can be formulated as follows (see
\cite{FB,AB} for more details and a complete proof). 

\noi {\bf Wedderburn's Theorem:} Let $\a$ be a semisimple 
(finite-dimensional) complex algebra with unit. Then $\a$ is a direct
sum of matrix algebras over $\Cop$
\be
\a = \bigoplus_{i} \M_{n_i}(\Cop) \,.
\ee

\noi {\it Sketch of Proof:} By semisimplicity $\a$ is a
direct sum of irreducible representations of $\a$, and because $\a$ is
finite-dimensional, this sum is finite
\be
\label{decomp}
\a=\bigoplus_{i=1}^{l} n_i\, \R_i\,.
\ee
Here $\R_i, i=1, \ldots l$ are pairwise inequivalent irreducible
representations of $\a$, and $n_i\in\Nop$ are the corresponding
multiplicities. We consider the vector space of endomorphisms
$\mbox{End}_{\a}(\a)$, which consists of all linear maps 
$\rho :\a \rightarrow \a$ for which $a\rho(b) = \rho(ab)$ for all 
$a,b\in\a$. This vector space has the structure of an algebra with
unit, and it is easy to see that
\be
\mbox{End}_{\a}(\a) \simeq \a^{op} \,,
\ee
where $\a^{op}$ is the algebra whose multiplication is defined by 
$(a \cdot b) \equiv (ba)$. (This identification relies on the fact
that $\a$ has a unit.) We therefore find
\begin{eqnarray}
\a^{op} & = & {\displaystyle
\bigoplus_{i=1}^{l} \mbox{End}_{\a}(n_i\, \R_i)}
\nonumber \\
& = & {\displaystyle
\bigoplus_{i=1}^{l} \M_{n_i}(\Cop) \,,}
\nonumber 
\end{eqnarray}
where we have used Schur's lemma in the last line. As
$(\a^{op})^{op}=\a$, we thus obtain the result, since
\be
(\M_{n}(\Cop))^{op} \simeq \M_{n}(\Cop)\,.
\ee
\smallskip

In our situation, we do not know a priori whether the algebra $\a$ has
a unit or not. We therefore have to analyze a more general case. 
First of all, because of semisimplicity of $\a$, we can still
decompose $\a$ as in (\ref{decomp}). If we assume that $\a$ is
finite-dimensional, then the direct sum is also necessarily finite. 

Next, we use the fact that a non-zero algebra necessarily has a unit
if it contains no non-zero nilpotent ideals 
\cite[Section 5, Theorem 26]{AB}. If $\a$ has no such ideals, we can
apply Wedderburn's theorem, and the structure of $\a$ is as described
above. Otherwise, suppose that $I$ is a non-zero nilpotent ideal, 
{\it i.e.} a subspace $I\subset \a$, such that $ac, cb \in I$
for every $a,b\in\a$, $c\in I$, and $c_1 \cdots c_n=0$ for some
(fixed) $n\in\Nop$ and arbitrary $c_i\in I$. As the representations
$\R_i$ are irreducible, every $c\in I$ either has to map $R_i$ onto
itself, or $c \R_i=0$. However, because of the nilpotency of $I$,
$c^n \R_i =0$, and the first possibility can be ruled out. This
implies that every element of $I$ annihilates every element in $\R_i$,
and thus, that $I$ acts trivially on $\a$. It is clear that the sum of
two nilpotent ideals is again a nilpotent ideal, and therefore
$\a$ contains a maximal nilpotent ideal; we shall denote this maximal
ideal by $I$.

Because of semisimplicity, $\a$ can be decomposed as 
$\a=I \oplus \hat{\a}$, where $\hat{\a}$ is a representation of
$\a$. By construction, $\hat\a$ does not have any non-zero nilpotent
ideals, and therefore possesses a unit. We can therefore apply
Wedderburn's theorem to $\hat{\a}$, to conclude that it is a direct
sum of matrix algebras.  

On the other hand, as every element of $I$ vanishes on any element of
$\a$, it follows that a basis $e_1, \ldots e_p$ can be chosen for $I$
so that 
\be
(e_i, e_j) = \delta_{ij} \qquad 
e_i e_j = 0 \qquad
e_i a = 0 
\ee
for every $a\in\hat{\a}$. This describes the structure of $\a$
completely.

\section{Traces over matrix algebras}
\setcounter{equation}{0}

Suppose that $\a$ is a complex (ungraded) matrix algebra, and that
there exists a bilinear form on $\a$ which satisfies the cyclicity
condition  (\ref{cyclic}). We want to show that 
\be
\langle a, b \rangle = \lambda\,  \tr (ab) \,,
\ee
where $\lambda\in\Cop$ is some normalization constant, and 
$\tr$ is the matrix trace in the defining realization of the algebra. 

We can choose a basis $E_{ij}$ for the algebra in terms of the
matrices whose only non-zero the entry is $1$ at the $(i,j)$ position; 
the defining relations of the algebra are then
\be
E_{ij} E_{kl} = \delta_{jk} E_{il} \,.
\ee
Because of the cyclicity of the bilinear form we have 
\be
\langle E_{ij}, E_{kl} E_{mn} \rangle  =  
\langle E_{kl}, E_{mn} E_{ij} \rangle \, ,
\ee
and therefore 
\be 
\label{identity}
\delta_{lm} \langle E_{ij}, E_{kn} \rangle  = 
\delta_{ni} \langle E_{kl}, E_{mj} \rangle\,. 
\ee
This then implies that
\be
\label{inner}
\langle E_{ij}, E_{kl} \rangle = \lambda_{ij} \delta_{jk} \delta_{il}
\,. 
\ee
Because of (\ref{inner}), (\ref{identity}) then implies that 
$\lambda_{ij} = \lambda_{jl}$. We therefore find that
\be
\lambda \equiv \lambda_{11} = \lambda_{1l} = \lambda_{lm} \,,
\ee
where we have used the identity twice. This completes the proof. 
 
It also follows from (\ref{inner}) that in order for the inner product
$(a,b)=\langle a^\ast, b \rangle$ to be positive definite, the 
star operation must coincide with hermitian conjugation of matrices.

\end{document}